\documentstyle[12pt]{article}

\newcommand{\be}{\begin{equation}}
\newcommand{\ee}{\end{equation}}

\newcommand{\ba}{\begin{eqnarray}}
\newcommand{\ea}{\end{eqnarray}}

\newcommand{\la}{\lambda}

\newcommand {\th}{\theta}

\newcommand{\Tr}{\rm Tr}
\newcommand{\tr}{\rm tr}

\newcommand{\e}{\epsilon}

\begin{document}

\hsize36truepc\vsize51truepc
\hoffset=-.4truein\voffset=-0.5truein
\setlength{\textheight}{8.5 in}

\begin{titlepage}
\begin{center}
\hfill \\
\hfill LPTENS-02/37\\
\vskip 0.6 in
{\large New correlation functions for random matrices
and integrals over supergroups}

\vskip .6 in
       \begin{center}
 {\bf E. Br\'ezin$^{a)}$}{\it and} {\bf S. Hikami$^{b)}$}
\end{center}
\vskip 5mm
\begin{center}
{$^{a)}$ Laboratoire de Physique
Th\'eorique, Ecole Normale Sup\'erieure}\\ {24 rue Lhomond 75231, Paris Cedex
05, France{\footnote{
 Unit\'e Mixte de Recherche 8549 du Centre National de la
Recherche
Scientifique et de l'\'Ecole Normale Sup\'erieure.
 } }}\\
 {$^{b)}$ Department of Pure and Applied Sciences,
}\\ {University of Tokyo,
Meguro-ku, Komaba, Tokyo 153, Japan}\\

\end{center}

 \vskip 0.5 cm
{\bf Abstract}
\end{center}
\vskip 14.5pt

The averages of ratios  of characteristic polynomials
${\rm det}(\lambda - X)$ of  $N \times N$ random matrices
$X$, are investigated in the large N limit for the GUE, GOE
and GSE ensemble.
The density of states and the two-point correlation function are derived
from these ratios.
The method relies on an extension of the
Harish-Chandra-Itzykson-Zuber  integrals to the GOE ensemble and to
supergroups , which  are explicitly evaluated as
 solutions
of heat kernel differential equations.
An  external matrix source, linearly coupled to the random matrices, may
also be added to the Gaussian distribution, and  allows for a discussion
of universality of the GOE results  in the large N limit.

\end{titlepage}
\setlength{\baselineskip}{1.5\baselineskip}
%\setcounter{page}{1}

%****************************************************************

\section{  Introduction }

 In this article we consider correlation functions involving
characteristic polynomials of $N\times N$ random matrices $X$ belonging to
the Gaussian unitary (GUE), Gaussian orthogonal ensemble (GOE) or
Gaussian symplectic ensemble (GSE), of the
following type
\be \label {ratio}F(\lambda_1,\cdots,
\lambda_k ; \mu_1\cdots  \mu_k) = \langle \prod_{\alpha=1}^k
\frac{{\rm{det}}(\lambda_{\alpha} -X)}{{\rm{det}}(\mu_{\alpha}
-X)}\rangle.\ee  The
reason for considering expectation values of such characteristic
polynomials are the following. First they turn out to be simpler than the
usual correlation functions of the type $\displaystyle\langle\prod_{\alpha
=1}^k
{\rm{Tr}}
\frac{1}{z_{\alpha}-X}\rangle$, but they may be used to recover the same
information. For instance
\be \frac{\partial}{\partial \lambda}
\langle\frac{{\rm{det}}(\lambda -X)}{{\rm{det}}(\mu
-X)}\rangle\vert_{\mu=\lambda} = \langle
{\rm{Tr}}\frac{1}{\lambda-X}\rangle\ee
and so on. Next   we know \cite{BH4,BH5,BH6} that  correlations
of  products
 $ < \prod_{\alpha=1}^k {\rm det}(\lambda_\alpha - X) >$ are universal in
the Dyson limit, in which the distance between the $\lambda$'s is of
order of the mean spacing \cite{Dyson1,Dyson2,Mehta}. This universality is
known to hold if the probability distribution is non-Gaussian, or if it
includes an external matrix source  linearly coupled to the random
matrices. Presumably this
extends to the ratios (\ref{ratio}) considered here and some evidence in
that direction will be presented below, when the Gaussian measure if
modified by the presence of an external source.

The calculations will be shown to
involve integrals over groups which go often beyond the case considered by
Harish-Chandra, Itzykson and Zuber (HIZ) \cite{Harish-Chandra,
Itzykson-Zuber}and the main point of this article is to analyze such
cases.  Such integrals appear at many places in random matrix theory.  For
instance they are essential  in the problem of a random Hamiltonian
$H$, an Hermitian $N\times N$ matrix, which is the sum of a non-random
$H_0$ and of a Gaussian random $V$ :
\be H= H_0+ V\ee
with \be P(V) = \frac{1}{Z} e^{-\frac{N}{2}{\rm{Tr}} V^2}.\ee
The probability law for the matrix elements of $H$ is thus
\be P(H) =\frac{1}{Z'} e^{-\frac{N}{2}{\rm{Tr}} H^2 + N
{\rm{Tr}}HH_0}.\ee
When $H_0$ is non-zero this measure is not invariant under a change of
basis, i.e. under the orthogonal, unitary or symplectic groups
appropriate to real symmetric, hermitian or quaternionic matrices.  If one
wants to write now the probability distribution for the eigenvalues of
$H$, one writes
$H= U\Lambda U^{-1}$ with $\Lambda$ diagonal, and $U$ a group element
appropriate to the ensemble there is a well-known Jacobian
\be dH =   dU\vert \Delta(\lambda_1,\cdots,
\lambda_N) \vert ^{\beta}\prod_1^N d\lambda_a,\ee
with $\beta=1,2$ or $4$. and the probability distribution for the
eigenvalues is given by
\be P(\lambda_1,\cdots,
\lambda_N) = \frac{1}{Z}\vert\Delta(\lambda_1,\cdots,
\lambda_N) \vert ^{\beta} e^{-\frac{N}{2}\sum\lambda_a^2}\int dU e^{N
{\rm{Tr}}( U\Lambda U^{-1}H_0)}.\ee
Therefore a group integration has to be performed.

Similar integrals occur in many places of random matrix theory. For
instance for chains of random matrices on a lattice with couplings of the
form $\exp {\rm{Tr}} H_nH_{n+1}$, clearly the joint probability for
eigenvalues involves a similar group integration over the relative
$U_n^{-1}U_{n+1}$.

 In the present work the  correlation functions
(\ref{ratio}) will be written as supersymmetric integrals and
diagonalization through supergroup transformations will lead us to
consider new types of HIZ group integrals.

  The literature on supersymmetric integration is very rich. Efetov
\cite{Efetov} has already obtained expressions of the correlations for
those same ratios (\ref{ratio})
  by supersymmetric techniques.
  In the unitary case, in which $X$ is a complex hermitian matrix,
    the calculation of the n-point correlation is done explicitely
    with the help of the HIZ integral,
     which
     has been extended also
    to the supermatrix case
    \cite{G,Alfaro}.
    However for real symmetrix matrices, for general $k$,
   the supermatrix formulation becomes complicated since
   the relevant HIZ integrals are no longer
explicit and simple. Recently, Guhr and Kohler \cite{GK1,GK2} have
studied those
integrals by the recursion formula of the size N of the matrix,
but it is a different approach which is proposed below. In a
previous work, we have considered such integrals for the symplectic group
\cite{BH3}, with the help  of differential equations satisfied by the
heat kernel for motions in the orthogonal group.
    This approach may be extended to include an external matrix source,
a modification for which the  standard
    orthogonal
    polynomial methods do not apply any more.

Since we have considered such HIZ integrals in the past, let us summarize
what was known earlier, beyond the standard HIZ integral for the GUE. In
the GOE case a mapping of the $N \times N$ integral for $\langle \prod_1^k
\det(\lambda_{\alpha}-X)\rangle$ to an integral over $k \times k$ matrices
invariant over the symplectic group was first derived. Then one found,
through the heat kernel differential equation, that the HIZ integral
was exactly given in that case by a semi-classical approximation
corrected by a {\it{finite}} number of terms. Those additional terms
are important, since they are  of order one in the Dyson limit ; once those
terms are known  one can determine this limit by a saddle-point
method. Here we return to those heat kernel equations for the supergroups
relevant to our present problem. It turns out that now the series of
correction terms to the semi-classical limit does not terminate. However in
the  scaling limit this series may be determined explicitly, and the problem
at hand is thus solved.

  The set-up of this article is as follows :  We first consider the
average of a single ratio
$\displaystyle
\langle\frac{{\rm det}(\lambda - X)}{{\rm det}(\mu - X)}\rangle $ for the GUE
and reduce it directly to quadratures, or alternatively we make use of
supersymmetric techniques with a simple version of the HIZ integral. Then the
same is done for the GOE. For a single ratio the  integration over the
supergroup variables is done through a generalization of the HIZ formula,
which is exactly given by the semi-classical approximation plus a finite
number of corrections. We then apply those supersymmetric techniques to
higher correlation functions. The required HIZ formula for supergroup
integration leads to a differential equation whose solution shows than
the semi-classical integration  has now to be corrected by an infinite
series. This infinite series originates  from the part of the supergroup
integration which comes from the orthogonal sub-group, and each term of
this series is of order one in the Dyson limit. However in the limit of
large matrices it turns out that one needs only to consider a special case
of those HIZ integrals, when the source matrices have only two distinct
eigenvalues. Then the whole series may be found in closed form and one
obtains explicitely the large N limit of the correlations functions for
ratios of characteristic polynomials. This is then generalized to k-point
functions and, continuing in $k$, one can study the zero-replica limit
$k\to 0$, which is an alternative way of deriving the usual correlation
functions from those determinental expectations. The Gaussian symplectic
ensemble (GSE) is discussed within the same appraoch. Those techniques
are  then extended further to probability measures involving an external
matrix source. This allows one to verify that the results in the  Dyson
limit are not modified by this source. Level spacing distributions are
then derived  within the same methods. The case of arbitrary $\beta$ is
then examined. Finally edge singularities are discussed as well.
%%%%%%%%%%%%%%%%%%%%%%%%%%%%%%%%%%%%%%%%%%%%%%%%%%%%%%%%%%%%%%%%%%%%%%%%%%%%

\section{Single ratio of characteristic polynomials in the GUE ensemble}

   Let us first consider a single ratio of  characteristic
polynomials,  defined as
  \be\label{2.1c}
  F_N(\lambda,\mu) = <\frac{{\rm det}(\lambda - X)}{{\rm det}(\mu - X)} >
  \ee
 The averages $< \cdots >$ are computed with respect to  the
Gaussian distribution
   \be\label{2.2c}
     P(X) = \frac{1}{Z}{\rm exp}( - \frac{N}{2} {\rm Tr} X^2 )
   \ee
     where the matrix $X$ is a $N \times N$ complex hermitian matrix.

    The  determinant ${\rm det}(\lambda - X)$ may be expressed as an
integral over
    Grassmann variables $\bar \theta$ and
    $\th$ \cite{BH3}.
   The determinant  ${\rm det}(\mu - X)$
in the denominator, may be expressed as  a Gaussian
   integral over commuting variables $z$ and $z^{*}$. Then
      \be\label{2.3c}
   F_N(\lambda,\mu) = \int \prod_{a=1}^N dz^{*}_a dz_a d\bar \th_a
d\th_a <  {\rm exp} i N
   [\bar \th_a (\lambda \delta_{ab}- X_{ab}) \th_b + z^{*}_a (\mu
\delta_{ab}-  X_{ab})z_b ] >
   \ee
in which $\mu$ is complex, with a small positive imaginary part. The
normalization  for Grassmann integration  which is used here is
\be \int d\th d\bar\th \   \th \bar\th = \frac{1}{\pi} \ee
We also adopt the convention
\be \overline{(\alpha \beta)} = \bar \alpha \bar \beta \ee
which implies
\be \bar {\bar \alpha} = -\alpha \ee
in order to maintain $ \bar \theta \theta$ invariant under the bar
operation.

    Then the Gaussian integration over $X$ is easily performed since

    \be\label{Yc}
    \int dX e^{- \frac{N}{2} \Tr X^2 + i N \Tr X Y} = e^{-\frac{N}{2} \Tr
Y^2 }.
    \ee
    with $Y_{ji}=  - \bar \th_i \th_j  - z^{*}_i z_j$.
    This gives
    \ba\label{2.7c}
    \Tr Y^2 &=& \Tr Y_{ij}Y_{ji}\nonumber\\
    &=& - (\bar \th_i \th_i)^2 + (z^{*}_i z_i)^2
    + 2(\bar \th_i z_i)(\th_j z^{*}_j)
    \ea

   One  then introduces  auxiliary commuting as well as  Grassmann
variables
     \ba\label{intc}
      &&\sqrt{\frac{N}{2\pi}}\int e^{- N\frac{b'}{2}^2 -N b' \bar \th_a
\th_a} db' = e^{\frac{N}{2} (\sum_a\bar
\th_a \th_a)^2},\nonumber\\
&&\sqrt{\frac{N}{2\pi}}\int e^{- N\frac{b}{2}^2 - iN b
z_a^*z_a } db = e^{- \frac{N}{2} (\sum_a z_a^* z_a)^2}
      \nonumber\\
      && \frac{\pi}{N}\int e^{- N \bar \eta \eta + N\eta (\bar \th z)
+ N\bar \eta (z^* \th)} d\eta d\bar  \eta =
      e^{- N (\bar \th z) (z^* \th)}
      \ea
       This allows to express $F_N$ as
    \ba\label{17}
    F_N(\lambda,\mu) &=& \frac{1}{2}\int db db'  d \eta d\bar\eta
  e^{-\frac{N}{2} (b^2 +
     b'^{2}
    + 2 \bar \eta \eta   )}\nonumber\\
    &&\times  [\int d z d z^{*} d  \th d\bar \th
    e^{ N ( i \lambda - b')\bar \th \th + i N (\mu -
    b)z^{*}z}
     e^{N \bar \eta \th z^{*} + N \eta \bar \th z }\ ]^N
    \ea
  Two strategies are possible at that stage :
either keeping both the
commuting and Grassmannian degrees of freedom, and use the supersymmetry
of the integral, or integrate out explicitely the Grassmannian variables.
They are equally simple for the single ratio of characteristic
polynomials considered here, however for higher correlation functions the
use of the invariance under supergroup transformations turn out to be
more powerful.\\
(i) {\underline{integration over the Grassmann variables}}\\
 Let us first give the expressions that one finds if one
first perform  the integrations over
$\bar\th,\th, z^*,z$. This yields
     \ba\label{cc2}
    F_N(\lambda,\mu) &=& \frac{1}{2}\int db db'   d \eta
d\bar\eta  e^{-\frac{N}{2} (b^2 +
     b'^{2} + 2 \bar \eta \eta )}
    \times  [\frac{ \lambda +ib'}{\mu - b}
+\frac{ \bar \eta \eta}{ (\mu-b)^2}]^N\nonumber\\
&=&(\frac{N}{2\pi})\int  [ \frac{(
\lambda +i b')^N}{ (\mu-b)^N}
   -\frac{ (\lambda +i b')^{N - 1}}{(\mu-b)^{N+1}}]e^{-\frac{N}{2} (b^2 +
     b'^{2}  )}
    db db' \ea

 It is convenient to use the (appropriately normalized) Hermite
polynomials
  \ba\label{Hermitec}
    H_n(\la) &=& e^{\la^2}(- \frac{d}{d\la} )^n e^{-
\la^2}\nonumber\\
    &=& \frac{2^n}{\sqrt{\pi}}\int_{-\infty}^{+\infty} (\la + i t)^n
e^{-t^2} dt
    \ea
which satisfy the recursion formula
\be H_n(\la) -2\la H_{n-1}(\la) +2(n-1) H_{n-2}(\la) = 0 \ee
 Then one obtains from (\ref{cc2})
\be\label{C3c}
   F_N(\lambda, \mu) =
\frac{1}{2^{N}\sqrt{\pi}}[H_{N}(\bar\la)\int_{-\infty}^{+\infty}
db
\frac{e^{-b^2}}{(\bar\mu-b)^N}
-
 H_{N-1}(\bar\la)\int_{-\infty}^{+\infty}
dt
\frac{e^{-b^2}}{(\bar\mu-b)^{N+1}}]
   \ee
in which $\mu$ has a small positive imaginary part and
 we have used for convenience
\be  \bar \la = \la \sqrt{\frac{ N}{2}} \hskip 2cm \bar \mu = \mu
\sqrt{\frac{ N}{2}}.\ee
After repeated integration by parts this gives
\be\label{C4}
   F_N(\lambda, \mu) =
\frac{1}{2^{N}\sqrt{\pi}(N-1)!}\int_{-\infty}^{+\infty}
db
\frac{e^{-b^2}}{(\bar\mu-b)} [H_{N}(\bar \lambda) H_{N-1}(b)-H_{N-1}(\bar
\lambda) H_{N}(b)]
   \ee  from which follows for an infinitesimal positive imaginary part of
$\mu$
\be  {\rm Im} F_N(\lambda,\mu) =
-\frac{\sqrt{\pi}}{2^{N}(N-1)!}
e^{-\bar \mu^2} [H_{N}(\bar \lambda)
H_{N-1}(\bar \mu)-H_{N-1}(\bar
\lambda) H_{N}(\bar \mu)] \ee
which vanishes as expected for $\lambda =\mu$ since $F_N(\mu,\mu) =1$.

The density of states $\rho(\lambda)$ is thus given by
   \be
   \rho(\lambda) = -\lim_{\mu \rightarrow \lambda} \frac{1}{\pi
N}\frac{\partial}{\partial \mu}
   {\rm Im} F_N(\lambda,\mu)
   \ee
and one recovers the well-known finite $N$ expressions \cite{Mehta}.\\
(ii) {\underline{use of supergroup symmetry}}\\
Let us return  to (\ref{17}) and use the supergroup
structure. One first write
\be -b'\bar\th\th -ibz^*z + \bar \eta \th z^* + \eta \bar \th z
= -i(z^*,\bar \theta)\  Q \left(\matrix{ z\cr
                        \th}\right) \ee

  with
\be Q = \left(\matrix{ b& i\bar\eta\cr
                    -i\eta&-i b'}\right)
      \ee
Then \be \frac{1}{2}{\rm Str} Q^2 = b^2+b'^2 +   2 \bar \eta \eta
 \ee
in which we have used the supertrace notation "Str" defined as follows.
    If one decomposes  the supermatrix $Q$ into four blocks
     \be
      Q = \left(\matrix{A&B\cr
                 C&D}\right),
     \ee
     in which  $A$  and $D$ have commuting matrix elements ,
whereas the matrix elements of
     $C$ and $B$  are anticommuting. Then by definition

 \be
       {\rm Str} Q = {\rm tr} A - {\rm tr} D
       .\ee
     Similarly one defines the superdeterminant
     \be\label{sdet}
     {\rm Sdet} [Q] = \frac{{\rm det}( A - B D^{-1} C)}{{\rm det} D} =
\frac{{\rm det}A}{{\rm det}(D - C A^{-1} B)}
     \ee
     The notation is justified by the fact that
$$\int d\bar \th d \th dz^*dz e^{-i\bar \phi Q \phi} =  ({\rm Sdet}
[Q])^{-1}$$ if $\phi = \left(\matrix{ z\cr
                        \th}\right)$ and $\bar\phi = (z^*,\bar\th)$).
If one defines the diagonal supermatrix
 \be \Lambda =\left(\matrix{ \mu &0 \cr 0 & \lambda}\right) \ee
     This leads to the compact expression for  $F_N(\lambda,\mu)$ as an
integral over
      the supermatrix $Q$,
      \ba \label{SIc}
      F_N(\lambda,\mu) &=&\frac{1}{2} \int dQ e^{-
\frac{N}{2} Str Q^2 }
\frac{1}{({\rm Sdet} (Q-\Lambda))^{N}}\nonumber\\
     &=& \frac{1}{2}e^{-N(\mu^2-\la^2)} \int dQ e^{-
\frac{N}{2} Str Q^2 - N Str Q
\Lambda}
\frac{1}{({\rm Sdet} Q)^{N}} .
      \ea

      One now performs a superunitary  transformation to diagonalize $Q$
\be \left(\matrix{z\cr
                    \theta}\right) = g \left(\matrix{z'\cr
                     \theta'}\right)\ee
with
\be g= \left(\matrix{a&\bar{\beta}\cr
                     \gamma & d}\right)\ee
in which $\beta$ and $\gamma$ are  anticommuting. Then, with the sign
convention used here  $(z^*,\bar{\theta})$ transform by
\be  g^{\dagger} = \left(\matrix{a^{*}&\bar \gamma\cr
                     \beta & d^{*}}\right).\ee
Unitary transformations $g^{\dagger} g= 1$ leave invariant the
quadratic form $z^*z + \bar{\theta}\theta$. One may diagonalize the
matrix $Q$ by such a unitary transformation
      \be
      Q = g \left(\matrix{u& 0\cr
                     0& it}\right) g^{\dagger}
     . \ee

 Then the integration measure on the
matrix $Q$ may be replaced by an integration over the eigenvalues
$u_1, t$ and a group integration, up to a Jacobian which is proportional
to $1/( u-it) ^2$.
     The representation (\ref{SIc}) requires then  to consider the
supergroup equivalent of the  HIZ integral defined by
     \be\label{IZc}
     I = \int dg e^{N {\rm Str} g Q g^{-1} \Lambda}
     \ee
    where $\Lambda$ and $Q$ are diagonal matrices.

  This integral yields
 \be I = \frac{N}{2\pi^2} (u - it)(\mu - \lambda) e^{ N( u \mu
- i  t \lambda)} .  \ee
The derivation of this formula relies on a formalism of differential equations
for $I$ which will be presented in detail in the coming sections.  Note that
the integral (\ref{IZc}) vanishes when the matrices
$\Lambda$ or $Q$ are multiple of the identity ; this is a consequence of the
Grassmannian degrees of freedom which imply that $\int dg \times 1 = 0$.

  Therefore we have obtained  the representation
  \be
  F_N(\lambda,\mu) = N(\lambda - \mu) \frac{e^{- N(\mu^2 - \lambda^2)}}{2
\pi^2}
  \int_{-\infty}^\infty \int_{-\infty}^\infty dt du
  ( \frac{i t}{u})^N \frac{1}{u - it} e^{- \frac{N}{2}(u^2 + t^2) - i N t
\lambda + N u \mu}
  \ee

   The density of states is then recovered through
   \ba
   \rho(\lambda) &=& - {\rm Im}\lim_{\mu\rightarrow \lambda}
   \frac{\partial}{\partial \mu}
   \int_{-\infty}^\infty \frac{dt}{2 \pi }\oint \frac{du}{2  \pi}
du(\frac{-it}{u})^N \frac{1}{u - i t}e^{- \frac{N}{2}(u^2 + t^2) - i N t
\lambda
+ N u \mu}
   \nonumber\\
   &=& \frac{1}{N}\int_{-\infty}^\infty \frac{dt}{2  \pi} \oint \frac{du}{2 i
\pi}
    (\frac{- it}{u})^N \frac{1}{u - i t}
   e^{- \frac{N}{2}(u^2 + t^2) - i N t \lambda + N u \lambda}
   \ea
   where the contour integral of $u$ circles around the
origin.
   This exact representation, valid for any $N$,  has been  derived
earlier by a completely  different method
\cite{BH1}, without supersymmetry. The semi-circle law is recovered
from there easily, in the large-N limit, by a saddle-point integration
(it is simpler for that to consider $\partial \rho/\partial \lambda$).

%****************************************

\section{Single ratio of characteristic polynomials in the GOE ensemble}

  We now consider a single ratio of  characteristic
polynomials of GOE,  defined as
  \be\label{2.3}
  F_N(\lambda,\mu) = <\frac{{\rm det}(\lambda - X)}{{\rm det}(\mu - X)} >
  \ee
 The averages $< \cdots >$ are computed with respect to  the
Gaussian distribution
   \be\label{2.4}
     P(X) =\frac{1}{Z} {\rm exp}( - \frac{N}{2} {\rm Tr} X^2 )
   \ee
     where the matrix $X$ is a $N \times N$ real symmetric matrix.

    As for the GUE, the ratio in (\ref{2.3}) is expressed by
      \be\label{2.5}
   F_N(\lambda,\mu) = \int \prod_{a=1}^N dz^{*}_a dz_a d\bar \th_a
d\th_a <  {\rm exp} i N
   [\bar \th_a (\lambda \delta_{ab}- X_{ab}) \th_b + z^{*}_a (\mu
\delta_{ab}-  X_{ab})z_b ] >
   \ee
in which $\mu$ is complex, with a small positive imaginary part.
We use the  normalization for
Grassmannian integration of the previous section.

    Then the Gaussian integration over $X$ is easily performed since

    \be\label{Y}
    \int dX e^{- \frac{N}{2} \Tr X^2 + i N \Tr X Y} = e^{-\frac{N}{4} \Tr
(Y^2 +
Y Y^T)}.
    \ee
    with $Y_{ji}=  - \bar \th_i \th_j  - z^{*}_i z_j$.
    This gives
    \ba\label{2.7}
    \Tr Y^2 &=& \Tr Y_{ij}Y_{ji}\nonumber\\
    &=& - (\bar \th_i \th_i)^2 + (z^{*}_i z_i)^2
    + 2(\bar \th_i z_i)(\th_j z^{*}_j )
    \ea
    \ba\label{2.8}
    \Tr Y Y^{T} = 2 (\bar \th_i z^{*}_i)(z_j \th_j) + (z^{*}_i
z^{*}_i)(z_j z_j)
    \ea
   One  then introduces  auxiliary commuting as well as  Grassmann
variables
     \ba\label{int}
      &&\sqrt{\frac{N}{\pi}}\int e^{- N{b'}^2 -N b' \bar \th_a \th_a} db'
= e^{\frac{N}{4} (\sum_a\bar
\th_a \th_a)^2},
      {\hskip 5mm} \sqrt{\frac{N}{\pi}}\int e^{- N{b}^2 - iN b z_a^*z_a }
db = e^{- \frac{N}{4} (\sum_a z_a^* z_a)^2}
      \nonumber\\
      &&\frac{N}{\pi}\int e^{- Nu^* u + \frac{iN}{2} u z_a^* z_a^* +
\frac{iN}{2} u^* z_a z_a} d u^* d u
      = e^{-\frac{N}{4} (z_a^* z_a^*)(z_b z_b)}\nonumber\\
      && \frac{\pi}{2N}\int e^{- 2N \bar \eta \eta + N\eta (\bar \th z)
+
N\bar \eta (z^* \th)} d
\bar \eta d \eta =
      e^{- \frac{N}{2} (\bar \th z) (z^* \th)}
\nonumber\\
      && \frac{\pi}{2N}\int e^{- 2 N\bar \eta' \eta' + N\eta' (\bar \th
z^*) +
N\bar \eta'(z \th)} d
\bar \eta' d \eta' =
      e^{- \frac{N}{2} (\bar \th z^*) (z \th)}
      \ea
       This allows to express $F_N$ as
    \ba\label{2.10}
    F_N(\lambda,\mu) &=& \frac{1}{4}\int db db' du du^*  d \bar\eta d\eta
d
\bar
\eta' d\eta'  e^{-N (b^2 +
     b'^{2} +u^*u
    + 2 \bar \eta \eta +  2 \bar \eta'\eta')}\nonumber\\
    &&\times  [\int d z d z^{*} d \bar \th d \th
    e^{ N ( i \lambda - b')\bar \th \th + i N (\mu -
    b)z^{*}z}\nonumber\\
    && e^{N \bar \eta \th z^{*} + N \eta \bar \th z + \frac{i}{2} N (
u^{*}
    z^{2} + u {z^{*}}^{2}) +
    N \eta' z^{*}\bar \th  + N \bar \eta' \th z}]^N
    \ea
  Two strategies are possible, as explained in the previous section.
    Let us first give the expressions that one finds if one
first perform  the integrations over
$\bar
\th,
\th, z^*,z$. Noting that
     \be\label{2.11}
     \int e^{i N (\mu - b) z^{*}z + \frac{i}{2} N (u^{*} z^{2}
     + u {z^{*}}^{2})}dz dz^{*} = \frac{i\pi}{N \sqrt{(\mu - b)^2
-  |u|^2}}
     \ee
     the  integration over $z,z^*, \bar\th, \th$, yields
     \ba\label{DTheta}
    F_N(\lambda,\mu) &=& \frac{1}{4}\int db db' du du^*  d \bar\eta
d\eta d\bar\eta' d\eta'  e^{-N (b^2 +
     b'^{2} +u^*u
    + 2 \bar \eta \eta +  2 \bar \eta'\eta')}
    \times  [\frac{ \lambda +ib'}{((\mu - b)^2 - |u|^2)^{1/2}}
\nonumber\\&&-\frac{1}{((\mu-b)^2-\vert u\vert^2)^{3/2}}\{u\bar \eta'
\eta + u^*
\bar
\eta
\eta' -(\mu-b) (\bar \eta \eta +\bar \eta'\eta')\}]^N
\ea

     One can next integrate over the remaining Grassmann variables
$\bar \eta, \eta, \bar \eta',\eta'$ and denoting
$D =
\sqrt{(\mu - b)^2 -  |u|^2}$, one obtains at the end

   \ba\label{C2}
   F_N(\lambda, \mu) &=& (\frac{N}{\pi})^2\int \frac{1}{D^N} [ (
\lambda +i b')^N -
   \frac{ 1 }{D^2} (\mu - b) (\lambda +i b')^{N - 1}
   + \frac{N - 1}{4 N D^2}( \lambda +ib')^{N - 2}]\nonumber\\
   &\times&e^{-N(b^2 + b'^2+ u^{*} u)} db db' d u d u^{*}
   \ea
 It is convenient to use the (appropriately normalized) Hermite
polynomials defined by (\ref{Hermitec}).
   Then, using a few integrations by parts, one obtains from (\ref{C2})
\ba\label{C3}
   &&F_N(\lambda, \mu) =
-\frac{(N-1)}{2^{N-1}\pi}H_{N-2}(\bar\la)\int_{-\infty}^{+\infty}
\frac{db}{\sqrt {\pi}}
\frac{e^{-b^2}}{(\bar\mu-b)^N}
\nonumber\\ &+&
\frac{1}{2^{N-1}\pi}H_{N-1}(\bar\la)\int_{-\infty}^{+\infty}
\frac{db}{\sqrt {\pi}}e^{-b^2 }(\bar\la -2b) \int_{0}^{\infty}d\rho
e^{-\rho}
\frac{1}{[(\bar\mu-b)^2-\rho]^{N/2}}
   \ea
in which $\mu$ has a small positive imaginary part and
 we have used for convenience
\be  \bar \la = \la \sqrt N \hskip 2cm \bar \mu = \mu \sqrt N.\ee

The
relation (\ref{C3})  for the average ratio allows one to find in
particular the density of states,
   and the density of states $\rho(\lambda)$ is thus
   \be
   \rho(\lambda) = - \lim_{\mu \rightarrow \lambda} \frac{1}{\pi
N}\frac{\partial}{\partial \mu}
   {\rm Im} F_N(\lambda,\mu)
   \ee
when ${\rm Im} \lambda \to 0$.

  The explicit formula (\ref{C3}) allows one to recover standard
results derived by the method of skew orthogonal polynomials
\cite{Mehta}.  In order to write explicitely the imaginary part of $F_N$
one has to distinguish between even and odd N . Let us for instance  work
with
$N=2M$ and check
(\ref{C3})
    for N=2 and N=4.
    Using
    \be\label{Im}
    {\rm Im}\int_{-\infty}^{+\infty} db \frac{e^{-b^2}}{(\bar\mu -
b)^N} =     - \frac{\pi}{(N-1)!} H_{N-1}(\bar\mu) e^{-\bar\mu^2}
    \ee
and
\be\label{Im2}
    {\rm Im}\int_{0}^{+\infty} d\rho
\frac{e^{-\rho}}{(\bar \mu-b)^2-\rho} =
-\pi\  {\rm sgn(\bar\mu-b)}
e^{-(\bar\mu-b)^2}\ee
    where ${\rm sgn(x)} = +1$ for $x>0$ and $-1$ for $x<0$.
    One thus finds, for $N=2$
    \be\label{density2}
   \sqrt{\pi} {\rm Im} F_2(\lambda,\mu) =
(\bar \mu-\bar \la)e^{- \bar\mu^2}- \bar \la( \bar \mu-\bar \la) B(\bar
\mu)
    \ee
in which
\be\label{B}
 B(x) = e^{-x^2/2}\int_0^x dy e^{-y^2/2}. \ee
    For $N=4$ we use ,
    \be
      \int_0^{\infty} \frac{1}{((\mu - b)^2 - \rho)^2} e^{-\rho} d\rho
      = \frac{1}{(\mu- b)^2} + \int_0^{\infty} \frac{1}{(\mu - b)^2 - \rho}
      e^{-\rho} d\rho
    \ee
  and obtain
 \be
    \sqrt{\pi}{\rm Im} F_4(\lambda,\mu) =
\frac{1}{16}e^{-\bar\mu^2} [H_2(\bar\la)H_3(\bar
\mu)-H_3(\bar\la)H_2(\bar \mu)] +\frac{1}{8}(\bar \mu-\bar
\la)H_3(\bar\la) (2\bar\mu e^{-\bar\mu^2} + B(\bar \mu))
    \ee
 We see that, as it should, ${\rm  Im} F_N(\mu,\mu)$ vanishes (since
$F_N(\mu,\mu)=1$).
    The above expressions agree with those derived from orthogonal
polynomials;
    \ba
    S_2(\lambda,\mu) &=&  \phi_0(\lambda)\phi_0(\mu) -
{\phi_0}^{\prime}(\lambda)
    \int_0^{\mu} \phi_0(t) dt \nonumber\\
    &=& \phi_0(\lambda)\phi_0(\mu) + \phi_1(\lambda)\phi_1(\mu) +
\phi_1(\lambda)
    \int_0^{\mu} \phi_2(t) dt.
    \ea
    where $\phi_n(\lambda) = (2^n n! \sqrt{\pi} )^{-1/2} e^{-\lambda^2/2}
H_n(\lambda)$.
    We find ${\rm Im} F_2= (\mu - \la) S_2(\lambda,\mu) e^{-\mu^2/2 +
\lambda^2/2}$.
    The $S_2(\lambda,\mu)$ is the diagonal part of a kernel, which is
a $2 \times 2$  quaternion matrix
    \cite{Mehta}.

    For general N, $S_N(\lambda,\mu)$ is given by
    \ba\label{SN}
    S_N(\lambda,\mu) &=& \sum_{i=0}^{N-1} \phi_i(\lambda) \phi_i(\mu) +
    (\frac{N}{2})^{1/2} \phi_{N-1}(\lambda) \int_{0}^{\mu}
    \phi_N(t) dt\nonumber\\
    &=& \sum_{i=0}^{\frac{N}{2}-1} \phi_{2 i}(\lambda)\phi_{2 i}(\mu) -
\sum_{i=0}^{\frac{N}{2}-1}
    { \phi'}_{2i}(\lambda)\int_{0}^{\mu}\phi_{2i}(t)dt
    \ea
    In the large N limit, the second term in (\ref{SN}) in the first line is
negligible compared
    to the first one. Indeed since $\phi_N(t)$ is a rapidly
oscillating
    function, its integral in the second term is asymptotically small.
Then one recovers  the semi-circle law in the limit
    $\lambda \rightarrow \mu$.
Therefore the first term of $S_N(\lambda,\mu)$
    is dominating ; it coincides with the
well-known universal kernel for the
GUE ensemble

    Let us further check for N=4 case. In this case, we obtain
    \ba
    F_4(\lambda,\mu) &=& \frac{1}{16} [ H_4(\lambda) - 2 \mu H_3(\lambda) + 6
H_2(\lambda)] B\nonumber\\
    &+& [ \frac{1}{8} \mu H_4(\lambda) + \frac{1}{8} (4 \mu^2 - 1)
H_3(\lambda) -
\frac{1}{2} \mu^3
    H_2(\lambda) ] e^{- \lambda^2}
    \ea
    Then we have
    \be
    \lim_{\lambda\rightarrow \mu} \frac{\sqrt{\pi}}{\lambda - \mu} {\rm
Im} F_4(\lambda,\mu)
    =- (\bar \mu^3 - \frac{3}{2}\bar\mu) B(\bar\mu) - ( 3 \bar\mu^2 +
\frac{3}{2}) e^{-\mu^2}
    \ee
    The right hand of this equation agrees precisely with
$S_4(\mu,\mu)$ in (\ref{SN})
    . This allows one to check also  the density of states since
    \be\label{dens}
     {\rm Im} {\lim_{\lambda\rightarrow \mu}}
      \frac{\partial}{\partial \mu}  F_N (\lambda,\mu) =
-S(\mu,\mu)
      \ee
     We have verified the known result of the density of state in the GOE,
     $\rho(\lambda) = \frac{1}{\pi N}S(\lambda,\lambda)$.

   We are interested in the  large N limit and the choice of variables
 $b,b'$ and
$\rho$
hereabove, is not the best one.
   Let us define instead $u_1$, $u_2$ and $t$, (analogous to the
variables
$u$ and $t$ which
have been used
      for the unitary case\cite{BH1,BH2} ; a comparison with
      the unitary case is very useful) as follows
     \be\label{tu}
       b= \frac{u_1 + u_2}{2}, \rho = \frac{(u_1 - u_2)^2}{4}
       \ee
     and $b' = t$.
The geometric meaning of those variables will become clear later when
the formalism of graded supermatrices is introduced.
     Then we have from (\ref{C2})
    \ba\label{csaddle}
    F_N(\lambda,\mu) &=&\frac{N^2}{2\pi^2}\int \int
\int_{-\infty}^{\infty} dt du_1 du_2
e^{- N(t + i \lambda)^2 - \frac{N}{2} ( u_1 + \mu)^2 -
\frac{N}{2}(u_2 + \mu)^2}\frac{(i
t)^N}{(u_1 u_2)^{N/2}} |u_1 - u_2|\nonumber\\
&\times&
      [ 1 + \frac{1}{2} \frac{u_1 + u_2}{i t (u_1 u_2)} + \frac{(N-1)}{4N}
\frac{1}{(it)^2 u_1 u_2}]
      \ea
     This representation is well adapted to the large-N limit. The
saddle points  are
     \ba\label{saddlepoint}
     t^{c} &=&   \frac{ - i \lambda \pm \sqrt{ 2 - \lambda^2}}{2} =
t_{+},t_{-}
     \nonumber\\
    { u_1}^{c} &=&   \frac{\mu \pm i \sqrt{ 2 - \mu^2}}{2} = u_{+}, u_{-}
\nonumber\\
    {u_2}^{c} &=&    \frac{\mu \pm i \sqrt{ 2 - \mu^2}}{2}
      .\ea

     We have analyzed before a similar large N behavior  for the kernel
of
     the unitary case \cite{BH9}.
     The factor  $(it/\sqrt{u_1 u_2})^N$ , is oscillatory in the large
N limit unless we  take
     the pair of  saddle points with $ (t,u_1,u_2) =
(t_{+},u_{+},u_{+})$
     and $(t_{-},u_{-},u_{-})$. The other six possibilities are
sub-leading.
     The fluctuations around the saddle-point
give
     \be\label{p1}
     \frac{1}{(f^{''}(t_c))^{1/2}}= \frac{1}{\sqrt{1 - \frac{1}{2 t^2}}}
     \ee
     and
     for $u_1$ and $u_2$ we have similar Gaussian fluctuations to take
into account, together with the
 factor $|u_1 - u_2|$. Then the Gaussian integration for the
u-fluctuations yields

     \be\label{p2}
     \frac{1}{(f^{''}(u_c))^{3/2}} = \frac{1}{(1 - \frac{1}{2 u^2})^{3/2}}
     \ee
     The second term in the integrand of (\ref{csaddle}) becomes at the saddle
point
     \be
     1 + \frac{1}{2} \frac{u_1 + u_2}{it (u_1 u_2)} + \frac{1}{4}
\frac{1}{(it)^2 u_1 u_2}
     = 1 + \frac{1}{it_c u_c} + \frac{1}{4 (it_c)^2 u_c^2}
     \ee
     which is $( 1 + \frac{1}{2 i t_c u_c})^2$. Noting that $u_c = i t_c$,
we find that this factor cancels the results
      (\ref{p1}) and
(\ref{p2}) of the Gaussian fluctuations.
     In the large N limit, then we have
     \be
    {\rm Im} F_N(\lambda,\mu) = \sin(N (\lambda - \mu)  \sqrt{2 - \lambda^2})
     \ee
     \ba
        \rho(\lambda) &=& \frac{1}{\pi N} \lim_{\mu\rightarrow \lambda}
\frac{\partial}{\partial \lambda}
        \sin[ N (\lambda - \mu) \sqrt{2 - \lambda^2}]\nonumber\\
        &=& \frac{1}{\pi}\sqrt{2 - \lambda^2}
        \ea
as expected.

%*********************************************************************
\section{Supermatrix diagonalization}

     We have obtained  an integral representation of the averaged ratio
of
     characteristic polynomials (\ref{DTheta}) by a Gaussian
integration over $\theta$'s and $z$'s.
     If, instead of integrating out the anticommuting variables,  we
keep bosonic and fermionic (supersymmetric)
degrees of freedom,
     the formulation becomes more transparent. Furthermore it turns out
that the diagonalization of the supermatrices by a supergroup element
provides an extremely useful representation in terms of the eigenvalues.

    Let us define the  super-matrix $Q$ by
     \be\label{Q1}
      Q = \left(\matrix{ b&  -u& i\bar \eta & i\eta'\cr
                         -u^*& b&i\bar\eta'& i\eta\cr
                         -i\eta& -i \eta'&-i b'&0\cr
                      i\bar\eta'&i \bar \eta&0&-i b'}\right)
      \ee
     which is such that one can write the various quantities
which appear in the integral representation (\ref{2.10}) as follows :
\ba  &&-ib z^*z+\frac{i}{2}  ( u^{*}
    z^{2} + u {z^{*}}^{2})  +\bar \eta \th z^*+ \eta \bar
\th z +
    \eta' z^{*}\bar \th  +  \bar \eta' \th z -b'\bar \th \th\nonumber
\\ &&=-\frac{i}{2}(z^*, z,\bar \theta,-\theta)\  Q \left(\matrix{ z\cr
                        z^*\cr
                         \th\cr
                      \bar \th}\right)\ea
and
\be \frac{1}{2}{\rm Str} Q^2 = b^2+b'^2 +  u^*u + 2 \bar \eta \eta + 2
\bar \eta' \eta' .\ee

    From the definition (\ref{sdet}) of  Sdet[Q] ,
     one has
       \be \label{SD}
{\rm Sdet} [Q] = -\frac{b^2-u^*u}{ [b' + \frac{i}{b^2-u^*u}\left(
b(\bar \eta \eta +\bar \eta'\eta') +u\bar \eta'\eta +u^* \bar
\eta \eta'\right)]^2}
       \ee
 Similarly we define the diagonal supermatrix $\Lambda$ , with
elements $(\mu,\mu,\la,\la)$ on the diagonal.
     This leads to a compact expression for  $F_N(\lambda,\mu)$ as an
integral over
      the supermatrix $Q$,
      \be \label{SI}
      F_N(\lambda,\mu) =\frac{1}{4}e^{N(\mu^2-\la^2)} \int dQ e^{-
\frac{N}{2} Str Q^2 - iN Str Q
\Lambda}
\frac{1}{({\rm Sdet} Q)^{N/2}}
     , \ee
     which is identical to (\ref{SIc}).

      Superunitary transformations on the variables are defined
by
\be \left(\matrix{z\cr
                    z^* \cr
                     \theta\cr
                    \bar{\theta}}\right) = g \left(\matrix{z'\cr
                    z'^* \cr
                     \theta'\cr
                    \bar{\theta'}}\right)\ee
with
\be g= \left(\matrix{a&b\cr
                     c & d}\right)\ee
in which $b$ and $c$ are $2\times 2$
matrices with anticommuting elements. Then
\be  g^{\dagger} = \left(\matrix{a^{\dagger}&\bar c\cr
                     -\bar b & d^{\dagger}}\right).\ee
Unitary transformations $g^{\dagger} g= 1$ leave invariant the
quadratic form $z^*z + \bar{\theta}\theta$. One may diagonalize the
matrix $Q$ by such a unitary transformation
      \be
      Q = g \left(\matrix{u_1& & &\cr
                     &u_2& & \cr
                     & & it &\cr
                     & & & it}\right) g^{\dagger}
      \ee
since one can see, through an identity similar to (\ref{SD}), that  it
has one doubly degenerate eigenvalue. Then the integration measure on the
matrix $Q$ may be replaced by an integration over the eigenvalues
$u_1,u_2,t$ and a group integration, up to a Jacobian which may be found
as follows. One considers a group element near the identity
\be g =  1 + i\epsilon\ee
with
 \be \epsilon = \left(\matrix{ia&  b\cr
                     \bar b& 0 }\right)\ee
with
\be  a = \left(\matrix{0 &v\cr
                     v^* &0}\right) .\ee

Starting from  a diagonal matrix $Q$ it is transformed under this
operation into $Q + \delta Q$, with
\be\label{Jacobian}
 \delta Q = i [\epsilon, Q ] = i \left(\matrix {0 & v(u_2-u_1) &
(it-u_1)b_{11}&
(it-u_1) b_{12}\cr
                          v^*(u_1-u_2)&0&(it-u_2)b_{21} &(it-u_2)b_{22}\cr
                          -(it-u_1)\bar b_{11}&
-(it-u_2)\bar b_{21}&0&0\cr -(it-u_1)\bar b_{12}&
-(it-u_2)\bar b_{22}&0&0}\right).\ee
    The Jacobian J is then simply
      \be\label{Jacobian2}
      J = \frac{|u_1 - u_2|}{(it - u_1)^2 (it - u_2)^2}
      \ee

     The representation (\ref{SI}) leads  to consider the supergroup
equivalent of the  HIZ integral defined by
     \be\label{IZ}
     I = \int dg e^{N {\rm Str} g Q g^{-1} \Lambda}
     \ee
    where $\Lambda$ and $Q$ are diagonal matrices, $\Lambda = {\rm
diag}(\mu,\mu,\lambda,\lambda)$
    and $Q = {\rm diag}(u_1,u_2,it,it)$.
    (We are still dealing  here with   a single ratio
(\ref{2.3}) of characteristic polynomials and we shall return to this
point for higher correlation functions).
  The HIZ integral $I$ may be obtained as the solution of
a heat kernel differential
  equation. Indeed  it satisfies the equation
   \be\label{L1}
   \Delta_Q I(Q) = \epsilon I(Q)
 \ee
in which $\Delta$ is  the Laplacian with respect to the matrix
elements of Q, sum of second derivatives with respect to the five
commuting variables, and to the four Grassmannian ones, and
\be \e = 2N^2(\mu^2-\lambda^2).\ee Since the
integral
$I$ is invariant under supergroup transformations of $Q$, it is a
function of the three distinct eigenvalues of Q
$u_1,u_2,it$ and one may express the Laplacian as a second order
differential operator with respect to those eigenvalues :
 \be\label{L2}
    \frac{1}{2J}  \frac{\partial}{\partial t} J  \frac{\partial}{\partial t}
I(Q)
  + \frac{1}{J} \sum_{\alpha=1}^2  \frac{\partial}{\partial u_\alpha} J
\frac{\partial}{\partial u_\alpha} I(Q)
  = -\epsilon I(Q)
  \ee
  (The factor 1/2 comes from the degeneracy of the eigenvalue $it$).
 Let $h$ denote the square root of the Jacobian
 \be\label{L3}
 h = J^{1/2}
 \ee
 and substitute in the differential equation  $I(Q) = \chi/h$.
 Then, with the help of the identity
 \ba\label{L4}
   &&\frac{1}{2 h} \frac{\partial}{\partial t}( h^2 \frac{\partial}{\partial
t}\frac{\chi}{h}) + \frac{1}{h} \sum_{\alpha=1}^2
   \frac{\partial}{\partial u_\alpha}( h^2 \frac{\partial}{\partial
u_\alpha}\frac{\chi}{h}) \nonumber\\
   &=& \frac{1}{2} \frac{\partial^2 \chi}{\partial {t}^ 2} +
\sum_{\alpha=1}^2
    \frac{\partial^2 \chi}{\partial { u_\alpha}^2} - \frac{\chi}{2 h}
   \frac{\partial^2 h}{\partial {t}^2} - \frac{\chi}{h} \sum_{\alpha=1}^2
   \frac{\partial^2 h}{\partial {u_\alpha}^2},
 \ea
the first derivative terms of
$\chi$
cancel in the differential equation and one obtains, up to contact terms
which have a vanishing contribution to the final result,
\be\label{L5}
    \frac{1}{2} \frac{\partial^2 \chi}{\partial {t}^2}  +
    \sum_{\alpha=1}^2 \frac{\partial^2 \chi}{\partial {u_\alpha}^2} +
     [
       - \sum_{\alpha=1}^2 \frac{1}{(i t - u_\alpha)^2}
    + \frac{1}{2}  \frac{1}{(u_1 - u_2)^2}]\chi
    = -\epsilon \chi
    \ee
If one substitutes
\be\label{L6}
\chi = e^{- 2 i N t \lambda + N u_1 \mu + N u_2 \mu} \sqrt{\vert u_1 -
u_2\vert  }\ g
\ee
(the square root factor comes from the degeneracies in the matrix
$\Lambda$),  the
  pole  $1/(u_1 - u_2)$ in (\ref{L5}) is cancelled.
  Then the  HIZ integral $I$ becomes
 \be\label{L7}
   I = e^{- 2 i N\lambda t + N(u_1 + u_2) \mu} (it - u_1)(it - u_2)
(\lambda -
\mu)^2 g
   \ee
  in which $g$ satisfies the Laplacian (heat kernel)
equation,
    \ba\label{L8}
   && - 2 i N\lambda \frac{\partial g}{\partial t} + 2 N\mu \frac{\partial
g}{\partial u_1}
    + 2 N \mu \frac{\partial g}{\partial u_2} + \frac{1}{2} \frac{\partial^2
g}{\partial t^2}
    +  \frac{\partial^2 g}{\partial {u_1}^2} + \frac{\partial^2 g}{\partial
{u_2}^2}
    + \frac{1}{u_1 - u_2} (\frac{\partial g}{\partial u_1} - \frac{\partial
g}{\partial u_2})\nonumber\\
    &+&  [ - \frac{1}{(i t - u_1)^2 } - \frac{1} {(i t - u_2)^2} ]g = 0,
    \ea

    Remarkably enough, the solution of this equation is simply,
    \be\label{L9}
    g = 1 - \frac{1}{2 N(i t - u_1) (\lambda - \mu)} - \frac{1}{2 N(i t -
u_2)(\lambda - \mu)}.
   \ee
    Collecting all factors  from (\ref{L7}), together with the Jacobian
( \ref{Jacobian2})
$F_N(\lambda,\mu)$ may then be  written as
    \ba\label{L10}
    F_N(\lambda,\mu) &=& \int \frac{(i t)^N}{(u_1 u_2)^{\frac{N}{2}}}
\frac{|u_1
- u_2|}{(i t - u_1) (i t - u_2)}
     (\lambda - \mu)^2 [ 1 - \frac{1}{2 N (\lambda - \mu)}(\frac{1}{i t -
u_1} +
\frac{1}{i t - u_2})]\nonumber\\
    &\times& e^{- 2 i N t \lambda  + N (u_1 + u_2)\mu  - N t^2 - \frac{N}{2}
(u_1^2 + u_2^2)} dt du_1 du_2
    \ea
   From there one derives the density of states
    \ba\label{L11}
    \rho(\lambda) &=&\frac{1}{\pi N}{\rm Im} \lim_{\mu\rightarrow \lambda}
    \frac{\partial}{\partial \lambda}F_N(\lambda,\mu)\nonumber\\
    &=& - \frac{1}{8 \pi^2 N} {\rm Im} \lim_{\mu\rightarrow \lambda}
    \int dt du \frac{(i t)^N}{(u_1 u_2)^{\frac{N}{2}} }
    \frac{|u_1 - u_2|}{(i t - u_1) (i t - u_2)}
     [ \frac{1}{it - u_1} + \frac{1}{i t - u_2}] \nonumber\\
     &\times& e^{- 2 i N t \lambda  + N(u_1 + u_2)\mu - N t^2 -
\frac{N}{2} (u_1^2 + u_2^2)}
     \ea

       This result has been also obtained recently  by Guhr and
Kohler\cite{GK2}.
     Since this expression is quite different from (\ref{csaddle}),
derived in the section two,
     it is worth showing that it does gives the same result for the
density of states.
     For simplicity let us consider the  N=2
     case.

     We return to the  variables defined in
(\ref{tu}) ; then $\rho(\lambda)$ reads
     \be
       \rho(\lambda) =
     \frac{1}{8\pi N}{\rm Im} \int \frac{(i t)^2}{b^2 - \rho}\frac{ ( i t -
b)}{((it)^2 -
2 it
b + b^2 - \rho)^2}
     e^{- N(t + i \lambda)^2 - N(b - \lambda)^2 - N\rho}
     dt db d\rho
     \ee
     The imaginary part is evaluated as in (\ref{Im}),
     \ba
     \rho &=& \frac{1}{8 \pi N} \int {\rm sgn(b)}\delta(\rho - b^2)\frac{
(it)^2(
i t -
b)}{((it)^2 -
2 it b + b^2 - \rho)^2}
     e^{- N(t + i \lambda)^2 - N(b - \lambda)^2 - N\rho}
     dt db d\rho \nonumber\\
     &=& \frac{1}{8 \pi N} \int {\rm{sgn(b)}} \frac{it - b}{(it - 2b)^2}
e^{- N
(t + i
\lambda)^2 - N b^2 - N (b - \lambda)^2} db dt
     \ea
     The factor $it-b$ in the integrand, is equivalent to taking a
derivative with respect to
$\lambda$ of the exponent.
     We also  introduce  an auxiliary  integral over a variable  $s$ to
express the denominator,
     \ba
     \rho &=& \frac{1}{2}\frac{\partial}{\partial \lambda}
     [ \int_{-\infty}^{\infty} dt \int_0^{\infty} db \int_0^{\infty}ds
       s  e^{-s (2b - it) - N b^2 - N (b-\lambda)^2 - N (t + i
\lambda)^2}\nonumber\\
      &-&  \int_{-\infty}^{\infty} dt \int_{-\infty}^{0} db \int_0^{\infty}ds
      s e^{s (2b - it) - Nb^2 - N(b-\lambda)^2 - N(t + i \lambda)^2}]
      \ea
      This difference of integrals is an even function of $\lambda$,
since it may be expressed as,
      \be
      \rho = \frac{1}{2}\frac{\partial}{\partial \lambda} [ f(\lambda) -
f(-\lambda) ]
      \ee
      with $f$, after integration over $t$, is
      \ba
      f(\lambda) &=& \int_0^{\infty} \int_0^{\infty} ds db s e^{- 2 N (b -
\frac{\lambda}{2} +
      \frac{s}{2N})^2 + \frac{s^2}{4 N} - \frac{N}{2}\lambda^2 }\nonumber\\
      &=& \int_0^{\infty} ds s e^{\frac{s^2}{4 N} - \frac{N}{2}\lambda^2}
      \int_{- \frac{\lambda}{2} + \frac{s}{2N}}^{\infty} db e^{- 2 b^2}
      \ea
      Therefore, one obtains
      \ba
      \frac{\partial}{\partial \lambda} f &=& \int_0^\infty s e^{-
\frac{1}{4N}
( s - 2 N \lambda)^2} ds\nonumber\\
        &=& 2 N e^{- N \lambda^2} + N \lambda \sqrt{4 N \pi} + 2 N \lambda
\int_{- 2 N \lambda}^0
        e^{- \frac{z^2}{4 N}} dz
       \ea
             Dropping the odd terms in $\lambda$, we recover
             the expression (\ref{density2}) for the $N=2$  density
of states.
      For general N, the same method applies. Taking the
imaginary part
      of $\frac{1}{(b^2 - \rho)^{(N/2)}}$, gives  derivatives of the
delta-function, and
       using integration by parts, we obtain the expression for the
density of states
%%%%%%%%%%%%%%%%%%%%%%%%%%%%%%%%%%%%%%%%%%%%%%%%%%%%%%%%

\section {Large N limit : the semi-circle law}

    We now consider the large N limit of the density of states, based
on the representation
    (\ref{L11}). In the large N limit,
    one may use the saddle point method for the integration over $t$ and
$u_1,u_2$.
    Since the denominators in (\ref{L11}) have double poles, we again change
variables to $b = \frac{u_1 + u_2}{2},
    r = \frac{(u_1 - u_2)^2}{4}$.
    Then the integral becomes
    \ba
    \rho(\lambda) &=& \lim_{\mu\rightarrow \lambda} {\rm Im}
    \int_{-\infty}^{\infty} dt \int_{-\infty}^{\infty} db \int_0^{\infty}
dr
    \frac{(it)^N}{(b^2 - r)^{N/2}}\frac{2 it - 2 b}{[(it)^2 - 2 b (it) +
b^2
- r]^2}
    \nonumber\\
    &\times&
    e^{- N(t + i\lambda)^2
    -N b^2 + 2 N b \mu - N r}
    \ea
    Integrating by parts over  $r$, one is led to
    \ba\label{rhoint}
    &&\rho(\lambda) = - \frac{1}{8 \pi^2}\lim_{\mu\rightarrow \lambda} {\rm Im}
\int_{-\infty}^{\infty} dt \int_{-\infty}^{\infty} db
     (\frac{it}{b})^N \frac{1}{it - b} e^{- N (t + i \lambda)^2 - N (b -
\mu)^2}\nonumber\\
     &+& \frac{ N}{4 \pi^2} \lim_{\mu\rightarrow \lambda} {\rm Im}
     \int_{-\infty}^{\infty}  \int_{-\infty}^{\infty}
\int_{-\infty}^{\infty}dt
du_1 du_2
     \frac{(it)^N}{(u_1 u_2)^{N/2}} \frac{|u_1 - u_2|}{it - u_1} (1 -
\frac{1}{2
u_1 u_2})
     \nonumber\\
     &\times& e^{- N (t + i \lambda)^2 - \frac{N}{2}(u_1 - \mu)^2 -
\frac{N}{2}(u_2 - \mu)^2}\nonumber\\
    \ea
    In the second term, we have used again the variables $u_1$ and $u_2$
instead
of $b$ and $\rho$.
    The saddle points are
    \be
    t_{\pm} = \frac{- i \lambda \pm \sqrt{2 - \lambda^2}}{2}
    \ee
    \be
    b_{\pm} = u_{\pm} = \frac{\mu \pm \sqrt{\mu^2 - 2}}{2}
    \ee
   The two leading saddle-points are, $(t_{+},b_{+})$ and
$(t_{-},b_{-})$ . Other choices, such as $(t_{+},b_{-})$
   give  an oscillatory  behavior, which can be neglected in the
large N limit.The first term in (\ref{rhoint}) is same as GUE.
 For the second term, one needs to evaluate the Gaussian fluctuation around
 the saddle point since there is a factor $|u_1 - u_2|$.
  One retains also the pair of saddle points
$(t_{+},u_{1+},u_{2+})$
    and $(t_{-},u_{1-},u_{2-})$.
    The fluctuation around the saddle points, give a factor
$1/\sqrt{f''(t_c)} = 1/\sqrt{1 + \frac{1}{2 {t^2}_c}} =
     \frac{\sqrt{t_c}}{(2 - \lambda^2)^{1/4}}$ for the integration over
    $t$. Noting that
    \be
      it_{+} - b_{+} = it_{+} - u_{+} = (\lambda - \mu)
(\frac{t_{+}}{\sqrt{2 -
\lambda^2}})
     \ee
     we find that the factors $1/\sqrt{f''(t_c)}$ cancel,
     and the second term becomes same form as the first term.
     We obtain
by adding  by the two saddle points, in the limit  $\lambda \simeq
\mu$,
     \be
     \rho(\lambda)  = \frac{1}{\pi N}\lim_{\mu\rightarrow \lambda}
\frac{\sin[ N (\lambda -
\mu)
      \sqrt{2 - \lambda^2}]}{\lambda - \mu} = \frac{1}{\pi}
\sqrt{2-\lambda^2}
     \ee
     which is indeed the semi-circle law in the large N limit for the
density of states.

%***************************************************
\section{Ratio of two characteristic polynomials and two point correlation
function: preliminaries}

    We now consider the ratio of two characteristic polynomials,
    \be
   F_N(\lambda_1,\lambda_2,\mu_1,\mu_2)
   = < \frac{{\rm det}(\lambda_1 - X){\rm det}(\lambda_2 - X)}{{\rm
det}(\mu_1 -
X)
   {\rm det}(\mu_2 - X)} >
   \ee
for the GOE ensemble.
     We introduce again Grassmann variables $\theta_1,\theta_2$ and
complex numbers
$z_1, z_2$ to represent these
    determinants.
    \ba
    && F_N(\lambda_1,\lambda_2,\mu_1,\mu_2) \nonumber\\
    &=& \int \prod_{a=1}^N \prod_{\alpha=1}^2 d{z^*}_{\alpha a}
d z_{\alpha
a} d\bar \theta_{\alpha a} d \theta_{\alpha a}\nonumber\\
   &\times& < e^{iN [ \bar \theta_{1a} (\lambda_1\delta_{ab} -
X_{ab})\theta_{1b} + \bar \theta_{2a}
     (\lambda_2 \delta_{ab} - X_{ab}) \theta_{2b} +
     z_{1a}^*(\mu_1\delta_{ab} - X_{ab}) z_{1b} + z_{2a}^* (\mu_2
\delta_{ab} - X_{ab}) z_{2b}]}>
     \ea
    We integrate out the random matrix $X$ by (\ref{Y}).
     Noting that ${\rm tr}Y^2$ and ${\rm tr} Y Y^T$ are now
     \be
       {\rm tr} Y^2 = - (\bar \theta_\alpha \theta_\beta)(\bar \theta_\beta
\theta_\alpha) +
(z_\alpha^* z_\beta)(z_\beta^* z_\alpha)
       + 2 (\bar \theta_\alpha z_\beta)(z_\beta^* \theta_\alpha)
       \ee
       \be
       {\rm tr} Y Y^T =   (\bar \theta_\alpha \bar \theta_\beta)(
\theta_\beta \theta_\alpha) +
(z_\alpha^* z_\beta^*)(z_\beta z_\alpha)
       + 2 (\bar \theta_\alpha z_\beta^*)(z_\beta \theta_\alpha),
       \ee
       where $\alpha,\beta = 1,2$, and $(\bar \theta_\alpha \theta_\beta) =
\sum_{a=1}^N
\bar \theta_{\alpha a} \theta_{\beta a}$.
       As for the one-point correlation function, one
introduces auxiliary fields, $b_1, b_2, b'_1, b'_2, ...$,
       and use the Gaussian integral representation of (\ref{int}).
Then, one
integrates out $\bar \theta, \theta, z^*, z$.
       The result is then cast into the form,

   \be
    F_N(\lambda_1,\lambda_2,\mu_1,\mu_2) = \int [ {\rm Sdet} Q]^{-\frac{N}{2}}
e^{- \frac{1}{2}{\rm Str} Q^2 +
    {\rm Str} Q \Lambda}
    \ee
    where $Q$ is an $8\times 8$ super-matrix.
    \be\label{Q2}
    Q = \left(\matrix{b_1& - v_1& -v_3& -v_4&i \bar \eta_1& i  \eta_2&i \bar
\eta_3&i  \eta_4\cr
         -v_1^*& b_1&- v_4^* &- v_3^*&i \bar \eta_2& i \eta_1&i \bar
\eta_4& i \eta_3\cr
         - v_3^*&-v_4&b_2&-v_2&i \bar \eta_5&i \eta_6&i \bar \eta_7&i \eta_8\cr
         -v_4^*&-v_3&-v_2^*&b_2& i \bar \eta_6&i \eta_5&i \bar \eta_8&i
\eta_7\cr
         - i \eta_1&- i \eta_2&- i \eta_5&-i \eta_6&-ib'_1&0&-w_1&-w_2\cr
         i \bar \eta_2 & i \bar \eta_1& i \bar \eta_6& i \bar \eta_5&0&
-ib'_1&-w_2^*&-w_1^*\cr
         - i \eta_3&-i \eta_4&-i\eta_7&-i\eta_8&w_1^*&w_2&-ib'_2&0\cr
         i\bar\eta_4& i \bar\eta_3&i\bar \eta_8& i \bar
\eta_7&w_2^*&w_1&0&-i b'_2}\right)
         \ee
Again one needs to diagonalize $Q$ by a super-group transformation, and
find the Jacobian for the representation in terms of eigenvalues. This
yields a factor
 $(t_1 - t_2)^4$ for the  $(b', w)$ block matrices of $Q$, as
discussed in a previous article\cite{BH3}. The Jacobian for the
Grassmannian part
is again obtained by linearizing the group transformation near identity
as in (\ref{Jacobian}).
Then, in terms of the eigenvalues of $Q$, we have
    \ba\label{FN2}
     F_N(\lambda_1,\lambda_2,\mu_1,\mu_2) &=& \int
\frac{[(it_1)(it_2)]^N}{(u_1
u_2 u_3 u_4)^{N/2}}
     \frac{\prod_{i<j}^4 | u_i  - u_j | (t_1 - t_2)^4}{ \prod_{i=1}^4(it_1 -
u_i)^2
     \prod_{i=1}^4(it_2 - u_i)^2}e^{- N \sum t_i^2 - N \sum_j u_j^2}\nonumber\\
     &\times& I \  d t_1 d t_2 du_1 du_2 du_3 du_4
    \ea
    where $I$ is  the HIZ integral
    \be
      I = \int dg e^{N {\rm Str}g Q g^{-1} \Lambda}
     \ee
     It is sufficient to take  diagonal matrices $Q = {\rm
diag}(u_1,u_2,u_3,u_4,it_1,it_1,it_2,it_2)$,
     and  $\Lambda$  :
     $\Lambda =
(\mu_1,\mu_1,\mu_2,\mu_2,\lambda_1,\lambda_1,\lambda_2,\lambda_2)$.
   Note that the eigenvalues of $\Lambda$ are doubly degenerate.
    This HIZ integral is again computed by a taking the Laplacian
with respect to  matrix elements,
   a generalization of what appeared in our
previous work
    \cite{BH3}.

 The Jacobian $J$ in this case is the generalization of (\ref{Jacobian2}),
which
is
 derived by (\ref{Jacobian}),
 \be\label{B1}
  J = \frac{\prod_{i<j}^4|u_i - u_j| (t_1 -
t_2)^4}{\prod_{i=1}^2\prod_{\alpha=1}^4
 (i t_i - u_\alpha)^2}.
 \ee
 The Laplacian equation is
 \be\label{B2}
  \frac{1}{2J} \sum_{i=1}^2 \frac{\partial}{\partial t_i} J
\frac{\partial}{\partial t_i} I(Q)
  + \frac{1}{J} \sum_{\alpha=1}^4  \frac{\partial}{\partial u_\alpha} J
\frac{\partial}{\partial u_\alpha} I(Q)
  = \epsilon I(Q)
  \ee
  The factor 1/2 comes from the degeneracy of $t$.
 We set again
 $h = J^{1/2}$
and substitute  $\psi = \chi/h$.
 Then, the same identity  (\ref{L4})  eliminates the first
derivatives
of $\chi$.
 %*************
  Since the eigenvalues of $\Lambda$ are degenerate,
  we factor out $\sqrt{(u_1 - u_2)(u_3 - u_4)}$, and
  we have the following HIZ integral,
  \ba\label{HIZint}
  I &=&   \frac{\prod_{i=1,2;\alpha=1,..,4}(it_i - u_{\alpha})
  (\lambda_1 - \mu_1)^2 (\lambda_1 - \mu_2)^2 (\lambda_2 - \mu_1)^2
(\lambda_2 -
\mu_2)^2}
  {(t_1 - t_2)^2 \prod_{i<j}\sqrt{u_i - u_j}(\lambda_1 - \lambda_2)^2
  (\mu_1 - \mu_2)^2}\nonumber\\
  &\times&\sqrt{(u_1 - u_2)(u_3 - u_4)} g
    e^{- 2 i N t_1 \lambda_1 - 2 i N t_2 \lambda_2 + N (u_1 + u_2)\mu_1
  + N(u_3 + u_4)\mu_2} + {\rm{(perm.)}}
  \ea
   where (perm.) means sum over permutations of the $\lambda_i$'s and
$\mu_i$'s. The differential equation becomes
 \ba\label{D1}
  &&- 2 i N \lambda_1 \frac{\partial g}{\partial t_1} - 2 i N \lambda_2
\frac{\partial g}{\partial t_2}
  + 2 N \mu_1 \frac{\partial g}{\partial u_1} + 2 N \mu_1 \frac{\partial
g}{\partial u_2}
  + 2 N \mu_2 \frac{\partial g}{\partial u_3} + 2 N \mu_2 \frac{\partial
g}{\partial u_4}\nonumber\\
  &+&
  \frac{1}{2}\sum_{i=1}^2 \frac{\partial^2 g}{\partial {t_i}^2}
  + \sum_{\alpha=1}^4 \frac{\partial^2 g}{\partial {u_\alpha}^2}
  + \frac{1}{u_1 - u_2} (\frac{\partial g}{\partial u_1} - \frac{\partial
g}{\partial u_2})
  + \frac{1}{u_3 - u_4} (\frac{\partial g}{\partial u_3} - \frac{\partial
g}{\partial u_4})\nonumber\\
  &+& g [ - \frac{2}{(t_1 - t_2)^2} - \sum_{i,\alpha} \frac{1}{(i t_i -
u_\alpha)^2}
  + \frac{1}{2(u_1 - u_3)^2} + \frac{1}{2(u_1 - u_4)^2} + \frac{1}{2(u_2 -
u_3)^2}\nonumber\\
  &+& \frac{1}{2(u_2 - u_4)^2}] = 0.
  \ea

 The solution to this equation may be obtained by expanding in
1/N ; it contains several factors
\ba\label{D2}
     g &=& [1 -  \frac{i}{N t_{12}}] [ 1+ \frac{1}{4N}(\frac{1}{u_{13}} +
\frac{1}{u_{14}}
+ \frac{1}{u_{23}} + \frac{1}{u_{24}}) + \cdots ]\nonumber\\
 &\times& [ 1 - \frac{1}{2N}(\sum_{i=1}^2 \sum_{\alpha=1}^4
\frac{1}{\tau_{i\alpha}}) \nonumber\\
&+& \frac{1}{4 N^2}((\frac{1}{\tau_{11}} + \frac{1}{\tau_{12}})(
\frac{1}{\tau_{13}} +
\frac{1}{\tau_{14}}) + (\frac{1}{\tau_{21}} + \frac{1}{\tau_{22}})(
\frac{1}{\tau_{23}} +
\frac{1}{\tau_{24}}) \nonumber\\
&+& (\frac{1}{\tau_{11}} + \frac{1}{\tau_{12}} + \frac{1}{\tau_{13}} +
\frac{1}{\tau_{14}})
(\frac{1}{\tau_{21}} + \frac{1}{\tau_{22}} + \frac{1}{\tau_{23}} +
\frac{1}{\tau_{24}}))
\nonumber\\
&-& \frac{1}{8 N^3}((\frac{1}{\tau_{11}} + \frac{1}{\tau_{12}} )
(\frac{1}{\tau_{13}} + \frac{1}{\tau_{14}})(\frac{1}{\tau_{21}} +
\frac{1}{\tau_{22}}  +
\frac{1}{\tau_{23}} + \frac{1}{\tau_{24}} )\nonumber\\
&+& (\frac{1}{\tau_{21}} + \frac{1}{\tau_{22}} )
(\frac{1}{\tau_{23}} + \frac{1}{\tau_{24}})(\frac{1}{\tau_{11}} +
\frac{1}{\tau_{12}}  +
\frac{1}{\tau_{13}} + \frac{1}{\tau_{14}} ) )\nonumber\\
&+& \frac{1}{16N^4}(\frac{1}{\tau_{11}} + \frac{1}{\tau_{12}}
)(\frac{1}{\tau_{13}} + \frac{1}{\tau_{14}} )
(\frac{1}{\tau_{21}} + \frac{1}{\tau_{22}} )(\frac{1}{\tau_{23}} +
\frac{1}{\tau_{24}} )]
\ea
where we define
\ba
t_{12} &=& (t_1 - t_2) (\lambda_1 - \lambda_2)\nonumber\\
u_{13} &=& (u_1 - u_3) (\mu_1 - \mu_2), \hskip5mm
u_{14} = (u_1 - u_4)(\mu_1 - \mu_2)\nonumber\\
u_{23} &=& (u_2 - u_3)(\mu_1 - \mu_2), \hskip5mm
u_{24} = (u_2 - u_4)(\mu_1 - \mu_2)\nonumber\\
\tau_{11} &=& (i t_1 - u_1)(\lambda_1 - \mu_1),
\hskip5mm
\tau_{13} = ( i t_1 - u_3)(\lambda_1 - \mu_2)\nonumber\\
\tau_{12} &=& (i t_1 - u_2)(\lambda_1 - \mu_1), \hskip5mm etc.
\ea

Note that all these variables $Nt, N\tau,Nu$ are of order one in
Dyson's scaling limit. The series in  $1/\tau$ terminate at order
$1/N^4$. However  the second
factor in (\ref{D2}) is an infinite series
in powers of $1/N$, and the scaling limit requires a determination of the
full series.  Remarkably enough a closed solution may be found
 in the scaling limit. This series turns out to be simply the
HIZ integral for the GOE ensemble, and it  will
be discussed in the next section before returning to the two-point function.
%****************************************

\section{Heat kernel equation for non-unitary Itzykson-Zuber integrals}

Let us consider the  integral,
 \be\label{A1}
   I = \int dg e^{N{\rm{Tr}}( g A g^{-1} B)}
 \ee
in which $g$ runs over the orthogonal group $O(2k)$, with the usual  Haar
measure of integration.  We do consider group integration in this
section, not supergroups . We shall also later generalize it to other
values of
$\beta$ with, as usual, $\beta=1,2,4$ for the  GOE,GUE and GSE ensembles
respectively.
 The eigenvalues of $A$ are denoted $(u_1, \cdots, u_{2k})$, and the
eigenvalues of
$B$ as
 $(\mu_1,\cdots,\mu_{2k})$.
 Considered as a function of the matrix $A$ the integral $I$ satisfies a
Laplacian equation,
 \be\label{A2}
   \Delta_A I = \epsilon I
 \ee
 where $\Delta_A$ is the Laplacian with respect to the matrix
elements of $A$, and
\be \epsilon  = N^2{\rm{Tr}}B^2 = N^2\sum_1^{2k} \mu_a^2.\ee
 Since $I$ is a function  of the
eigenvalues
$u_i$ of $A$ one may replaced the Laplacian on the matrix elements by a
differential operator on the eigenvalues
 \be\label{A3}
    \Delta = \frac{1}{J} \sum_{i=1}^{2k}\frac{\partial}{\partial u_i} J
\frac{\partial}{\partial u_i}
 \ee
 where the measure $J$ is given by the absolute value of the Vandermonde
determinant,
 \be\label{A4}
 J = \prod_{i>j}^{2k} |u_i - u_j|
 \ee
 Let us investigate the simplest   case of a $2\times2$ matrix ($k=1$),
for which one obtains
 \be\label{A5}
 [ \frac{\partial^2}{\partial {u_1}^2} + \frac{\partial^2}{\partial {u_2}^2} +
 \frac{1}{u_1 - u_2} \frac{\partial}{\partial u_1} - \frac{1}{u_1 - u_2}
\frac{\partial}{\partial u_2}] I = \epsilon I
 \ee
 To eliminate the first order derivatives, one substitutes
 \be\label{A6}
   I = \frac{1}{\sqrt{u_1 - u_2}} \chi
 \ee
 Then (\ref{A2}) becomes
 \be\label{A7}
 \frac{\partial^2 \chi}{\partial {u_1}^2} + \frac{\partial^2 \chi}{\partial
{u_2}^2} + \frac{1}{2}
 \frac{1}{(u_1 - u_2)^2} \chi = \epsilon \chi
 \ee
 One then substitutes further
 \be\label{A8}
 \chi = e^{N u_1 \mu_1 + N u_2 \mu_2} f
 \ee
 Noting that $\epsilon = N^2 ({\mu_1}^2 + {\mu_2}^2)$, we obtain
 \be\label{A88}
 2 N \mu_1 \frac{\partial f}{\partial u_1} +  2 N \mu_2 \frac{\partial
f}{\partial u_2}
 + \frac{\partial^2 f}{\partial {u_1}^2} + \frac{\partial^2 f}{\partial
{u_2}^2}
+ \frac{1}{2 (u_1 - u_2)^2} f = 0.
 \ee
 From this differential equation one generates  the expansion of the
function
$f$  in powers of $1/N$
 \be\label{A9}
 f = 1 + \frac{1}{4 N(u_1 - u_2)(\mu_1 - \mu_2)} + \frac{9}{32 N^2 (u_1 -
u_2)^2
(\mu_1 - \mu_2)^2} + O(\frac{1}{N^3})
 \ee

 This series is in fact the expansion of a modified Bessel function, as
may be recognized on the differential equation. Indeed if we look for a
solution $f$, which is a function of the single scaling variable
\be x = N(u_1-u_2) (\mu_1-\mu_2)\ee
one sees that $f(x)$ should be the
solution of the differential equation
\be f''(x) + f'(x) +\frac{1}{4x^2} f = 0 .\ee
Therefore the $k=1$ solution is
 \be\label{A10}
 I = e^{\frac{N}{2}(u_1 + u_2)(\mu_1 + \mu_2)} I_0(\frac{N}{2}(u_1 -
u_2)(\mu_1 - \mu_2))
 \ee
 where $I_0(z)$ is a modified Bessel function, whose asymptotic
expansion is
 \be\label{A11}
 I_0(x) = \frac{e^{x}}{\sqrt{2\pi x}}[ 1 + \frac{1}{8 x} + \frac{9}{2 (8
x)^2} +
\frac{225}{6 (8 x)^3} +  \cdots ]
 \ee
 in agreement with (\ref{A9}) .

  In the expansion (\ref{A9}), we have implicitely assumed that $(u_1 -
u_2)$ is of order one in the Dyson limit, where $N(\mu_1 - \mu_2)$
  is of order one. However if  $(u_1 - u_2) \sim O(1/N)$, the previous
expansion is not valid and one has to perform a different expansion of the
solution of
  the differential equation  (\ref{A88}). Instead of (\ref{A9}) one
expands
  \be\label{A11a}
  f = C \sqrt{\vert u_1 - u_2\vert }  [ 1 - \frac{N}{2} (\mu_1 -
\mu_2)(u_1 - u_2) +
\cdots ]
  \ee
   in order to be able to deal with the regime  in which
the difference of the two arguments
$u_1$ and $u_2$ becomes
  order of $O(1/N)$. It corresponds of course
  to the small $x$-expansion of the modified Bessel function
(\ref{A11}).

Let us now proceed to the k=2 case, which is relevant to our
problem.
 We have $J = \prod_{1\leq i<j\leq 4} |u_i - u_j|$, and consider the
region
$u_1 > u_2 > u_3 >
u_4$. The differential equation for $I$ reads

 \be\label{A12}
 \sum_{i=1}^4 \frac{\partial^2}{\partial { u_i}^2}I + \sum_{i=1}^4
\sum_{j\neq i} \frac{1}{u_i - u_j}
 \frac{\partial}{\partial u_i}I = \epsilon I
 \ee
 Substituting  $I = \frac{1}{\sqrt{\vert\Delta(u)\vert}} \chi$,
 where $\Delta(u)$ is the  Vandermonde determinant of the $u_i$
(i=1,...4), one has
 \be\label{A14}
   \sum_{i=1}^4 \frac{\partial^2}{\partial {u_i}^2} \chi + \frac{1}{2} \chi
\sum_{i<j} \frac{1}{(u_i - u_j)^2} = \epsilon \chi
 \ee
  Again one substitutes
  $\chi = e^{N u_1 \mu_1 + N u_2 \mu_2 + N u_3 \mu_3 + N u_4 \mu_4} f$,
 and finds
 \be\label{A15}
  2 N \sum_{i=1}^4 \mu_i \frac{\partial f}{ \partial u_i} +
\sum_{i=1}^4\frac{\partial^2 f}{\partial {u_i}^2}
   + \frac{1}{2} \sum_{i<j}\frac{1}{(u_i - u_j)^2} f = 0
 .\ee
This equation has beautiful properties in the scaling limit of interest.
Before discussing this property,let us write the generalization of this
equation for arbitrary
$\beta$ ($\beta=1,2,4$ are the standard values for GOE, GUE and GSE
respectively), and for k-variables instead of four :
\be\label{A1500}
  2 N \sum_{i=1}^k \mu_i \frac{\partial f}{ \partial u_i} +
\sum_{i=1}^k\frac{\partial^2 f}{\partial {u_i}^2}
   -y \sum_{i<j}\frac{1}{(u_i - u_j)^2} f = 0
 ,\ee
in which $y$ stands for
\be y = \beta( \frac{\beta}{2}-1).\ee
It is a remarkable property that this equation has a solution which is a
function of the scaling variables
\be \tau_{ij} = N (\mu_i-\mu_j) ( u_i-u_j) \ee
alone, and not separately of the $u$'s and the $\mu$'s. Given the origin
of this equation, namely the HIZ integrals, it is clear that the solution
involves only dimensionless products of the type $\mu_i\cdot u_j$, and
that it is unchanged by a simultaneous permutation of $u_i$ and $u_j$
accompanied by the same permutation on $\mu_i\leftrightarrow\mu_j$. However a
direct proof or verification on the equation itself, turns out to lead to very
elaborate combinatorial identities  if one tries, for instance, to expand  the
solution  for large
$\tau$'s. The lowest orders are simple , but  the calculations become
very tedious, and non trivial,  at higher orders. Let us reproduce here
simply the  first terms :
 \be\label{A16}
  f = 1 -\sum_{i<j} \frac{y}{2 \tau_{ij}} -
  \sum_{i<j}\frac{y}{2}\frac{1}{ \tau_{ij}^2}
+ \frac{y^2}{8}(\sum_{i<j} \frac{1}{\tau_{ij}})^2 + \cdots
  \ee
 This expansion shows our point : the coefficients  are pure
numbers, completely independent of the parameters $\mu$'s, once the function is
expressed in terms of the $\tau_{ij}$'s.

This is a generalization of the
expansion of the modified Bessel function(\ref{A9}) ; one  recovers the
simple Bessel limit if one lets all the
$u$'s, except the first two,go to infinity.

    As  discussed hereabove, all we need in the large N limit,  is   a
solution  for the degenerate case,
$\mu_1 = \mu_2= \mu$,
    and $\mu_3 = \mu_4=\mu'$.
     To deal with this degenerate case, one writes  $f$ as
    \be\label{degenerate}
    f = \sqrt{\vert (u_1 - u_2) (u_3 - u_4)\vert } g
    \ee

    From (\ref{A15}), we have
    \ba\label{A17}
     &&2 N \mu  \frac{\partial g}{ \partial u_1} +  2 N \mu
\frac{\partial
g}{
\partial u_2}
     + 2 N \mu'  \frac{\partial g}{ \partial u_3} +  2 N \mu'
\frac{\partial
g}{
\partial u_4}
     + \sum_{\alpha=1}^4  \frac{\partial^2 g}{\partial {u_\alpha}^2}
\nonumber\\
     &+&
\frac{1}{u_1 - u_2}(\frac{\partial g}{\partial u_1} - \frac{\partial
g}{\partial
u_2})
+ \frac{1}{u_3 - u_4}(\frac{\partial g}{\partial u_3} - \frac{\partial
g}{\partial u_4}) \nonumber\\
     &+& [ \frac{1}{2} \frac{1}{(u_1 - u_3)^2}
     + \frac{1}{2} \frac{1}{(u_1 - u_4)^2}+ \frac{1}{2} \frac{1}{(u_2 -
u_3)^2}+
\frac{1}{2} \frac{1}{(u_2 - u_4)^2}] g= 0
    \ea

  This function $g$ may then be  obtained as an expansion in powers of
$\frac{1}{N}$. Terms involving  poles in
$\frac{1}{u_1 - u_2}$ and $\frac{1}{u_3 - u_4}$ do not appear in the
expression for
$g$. We have
 \ba\label{A18}
  g &=& 1 + \frac{1}{4 N (u_1 - u_3) (\mu - \mu')} +
   \frac{1}{4 N(u_1 - u_4)(\mu - \mu')} + \frac{1}{4 N (u_2 - u_3) (\mu -
\mu')}\nonumber\\
  &+& \frac{1}{4 N(u_2 - u_4) (\mu - \mu')} \nonumber\\
&+& \frac{9}{32 N^2(\mu - \mu')^2}[\frac{1}{(u_1 - u_3)^2} +
 \frac{1}{(u_1 - u_4)^2} + \frac{1}{(u_2 - u_3)^2} + \frac{1}{(u_2 - u_4)^2}
]\nonumber\\
&+& \frac{1}{16 N^2(\mu - \mu')^2} [ \frac{1}{(u_1 - u_3)(u_2 - u_4)}
+ \frac{1}{(u_1 - u_4)(u_2 - u_3)}]
\nonumber\\
&+& \frac{3}{16 N^2(\mu - \mu')^2}[\frac{1}{(u_1 - u_3)(u_1 - u_4)}
+ \frac{1}{(u_1 - u_3)(u_2 - u_3)}
+ \frac{1}{(u_1 - u_4)(u_2 - u_4)} \nonumber\\
 &+& \frac{1}{(u_2 - u_3)(u_2 - u_4)}]
+ O(\frac{1}{N^3})
  \ea

This expression has  a complicated structure ; one may write the
various terms as diagrams, but the coefficients depend upon the
topological character of those diagrams. For instance, the coefficients
at order $1/N^2$ the three types of diagrams have weights 9/32, 1/16 and
3/16. However, we find that in the large N limit, the expressions
simplify.

In the large N limit, if we concentrate on the saddle points $u_1 = u_2 =
u_{+}$ and
$u_3 =
u_4 = u_{-}$,
one obtains
\be\label{A21}
  g = 1 + \frac{1}{N (u_{+} - u_{-})(\mu - \mu')} + \frac{2}{N^2 (u_{+} -
u_{-})^2(\mu - \mu')^2} + O(\frac{1}{N^3}).
\ee
If we introduce  the scaling variable
 \be \label {A100} x = N (u_{+} - u_{-})(\mu - \mu'),\ee
one finds that $g(x)$ satisfies the  differential equation,
\be
  \frac{\partial^2 g}{\partial x^2} + ( 1 - \frac{1}{x} )\frac{\partial
g}{\partial x} +
\frac{g}{x^2} = 0
\ee
The expansion of $g$ in powers of $1/x$ is
\be
g = \sum_{n=0}^\infty \frac{n!}{x^n}
\ee

It is interesting to generalize this formula to general $k$, $(k > 4)$.
Denoting  $v_j = u_{k + j}$,(j=1,...k), one deals with the differential
equation,
 \ba\label{A25}
     &&2 N \mu \sum_{\alpha=1}^k \frac{\partial g}{ \partial u_\alpha}
 + 2 N \mu'\sum_{\alpha=1}^k \frac{\partial g}{ \partial v_\alpha} +
      \sum_{\alpha=1}^k  \frac{\partial^2 g}{\partial {u_\alpha}^2}
+ \sum_{\alpha=1}^k  \frac{\partial^2 g}{\partial {v_\alpha}^2}
\nonumber\\
     &+&
\sum_{\alpha < \beta}\frac{1}{u_\alpha - u_\beta}(\frac{\partial g}{\partial
u_\alpha}
- \frac{\partial
g}{\partial
u_\beta})
+ \sum_{\alpha < \beta} \frac{1}{v_\alpha - v_\beta}(\frac{\partial
g}{\partial
v_\alpha}
 - \frac{\partial
g}{\partial v_\beta}) \nonumber\\
     &+& g \sum_{\alpha=1}^k \sum_{\beta=1}^k [ \frac{1}{2}
\frac{1}{(u_\alpha -
v_\beta)^2}]
= 0
        \ea

For general $k$, the situation is similar.  In the scaling limit, in which
the leading saddle points involve only two distinct values  of the form
$u_a= u_+$ and
$v_a = u_-$,
 one introduces the same scaling variable $x$ (\ref{A100}), and
obtain  the differential equation,
\be\label{greplica}
  \frac{\partial^2 g}{\partial x^2} + ( 1 - \frac{k - 1}{x} )\frac{\partial
g}{\partial x} +
\frac{k^2}{4 x^2} g= 0.
\ee
An expansion in powers of $1/x$ of the solution follows easily :
\be\label{greplica2}
g = 1 + \sum_{p=1}^\infty \frac{[k ( k + 2)( k + 4) \cdots ( k + 2 ( p -
1))]^2}{2^{2 p} p! x^p}
\ee
One recovers the previous results for $k=1$ and $k = 2$ from
this expression.

We have considered hereabove  the GOE measure $\displaystyle J = \prod_{i>j} |
u_i - u_j|$ . Let us now go to arbitrary $\beta$, i.e. work with
$\displaystyle J =
\prod_{i>j} | u_i - u_j|^{\beta}$ ;  the GUE integral corresponds to  $\beta=2$
and the GSE to
$\beta=4$. When the eigenvalues $\mu_i$, which are set at the saddle-points in
the scaling limit,  degenerate again into two groups, one equal to
$\mu$ and the other one to
$\mu'$, one finds through identical steps
\ba\label{A27}
     &&2 N \mu\sum_{\alpha=1}^k \frac{\partial g}{ \partial u_\alpha}
 + 2 N \mu' \sum_{\alpha=1}^k \frac{\partial g}{ \partial v_\alpha} +
     + \sum_{\alpha=1}^k  \frac{\partial^2 g}{\partial {u_\alpha}^2}
+ \sum_{\alpha=1}^k  \frac{\partial^2 g}{\partial {v_\alpha}^2}
\nonumber\\
     &+&
\beta\sum_{\alpha < \beta}\frac{1}{u_\alpha - u_\beta}(\frac{\partial
g}{\partial  u_\alpha}
- \frac{\partial
g}{\partial
u_\beta})
+ \beta \sum_{\alpha < \beta} \frac{1}{v_\alpha -
v_\beta}(\frac{\partial  g}{\partial v_\alpha}
 - \frac{\partial
g}{\partial v_\beta}) \nonumber\\
     &+& \frac{\beta(2-\beta)}{2}[ \sum_{\alpha=1}^k \sum_{\beta=1}^k
\frac{1}{(u_\alpha - v_\beta)^2}] g
= 0 .
        \ea
In the scaling limit, in which the $u$'s and the $v$'s approach repectively
$u_+$ and $u_-$, one introduces again the scaling variable (\ref{A100}) and
one obtains
 the differential equation
\be
  \frac{\partial^2 g}{\partial x^2} + ( 1 - \frac{k - 1}{x} )\frac{\partial
g}{\partial x} +
\frac{k^2 \beta (2 - \beta)}{4 x^2} g= 0
\ee
which generalizes the previous GOE expression for $\beta = 1$. The expansion
of the general $\beta$, arbitrary $k$ solution, in powers of
$1/x$ follows
\be\label{A29}
g = 1 + \sum_{p=1}^\infty
 \frac{(\beta k)((2 - \beta)k)(\beta k + 2)( 2 + k(2 - \beta)) \cdots
( 2 ( p - 1) + k (2 - \beta))}{2^{2 p} p! x^p}
\ee
In  the unitary case, $\beta=2$, the solution reduces to the first term $g =
1$  : this is of course  the well-known Itzykson-Zuber result  which is
semi-classically exact.

 For $\beta=4$, and
$k=1$, we also see that the asymptotic expansion (\ref{A29}) stops at
first order, since the terms with $p>1$ vanish in the series  (\ref{A29}), a
fact that we had already found and   used
 in (\ref{D2}) for integrating over the $t_i$'s.
For $\beta = 4$, and $k=2$, it stops at third order. The fact that the
Itzykson-Zuber integral for this scaling limit of the GSE, is semi-classical
with only a {\it{finite}} number of corrections
 had already been discussed and used  in  \cite{BH3}.

Therefore one sees that, in the scaling limit of interest,  in which we
deal with an  Itzykson-Zuber integral for a degenerate case, there is a
remarkably simple expression for arbitrary  $\beta$ and arbitrary
dimension
$k$ of the group integration. The expression is either semi-classically exact
for the GUE, corrected by a finite number of terms for the GSE, or an
infinite, but explicit, series for the GOE. The fact that $k$ , the
dimension of the integral, appears as a parameter will allow us later to
continue in $k$ and use  the replica method in the
$k$ goes to  zero limit.

Returning to the GOE, let us note that if we had
  included an $i$ in the exponent of the
Itzykson-Zuber  integral
in (\ref{A1}), and multiplied by the factor $e^{i x}/x$, we would have
obtained for $k=2$
\ba\label{A101}
{\rm Re}[ \frac{e^{ix}}{x} g(i x)] &=& \frac{\cos x}{x} + \frac{\sin x}{x^2} -
\frac{2 \cos x}{x^3}
- \frac{6 \sin x}{x^4} + \cdots \nonumber\\
&=& \sum_{k=1}^\infty \frac{(k - 1)! \sin(x + \frac{\pi}{2}
k)(-1)^{(k-1)}}{x^k}
\ea
which is the  large $x$ asymptotic expansion   of the integral
\be
I = \int_x^\infty \frac{\sin z}{z} dz
.\ee
This integral  appears in the scaling limit of the resolvent-resolvent
correlation function fo the GOE ensemble.

%*********************************************
\section{Two point correlation function}

 We are now in position to return to the two point correlation function of the
GOE, which will be deduced from the ratio of products ot two characteristic
polynomials. Indeed one may obtain the resolvent-resolvent correlation function

\be\label{6.1}
\rho(\lambda_1,\lambda_2) =\frac{1}{\pi^2 N^2} \frac{\partial}{\partial
\lambda_
2}\frac{\partial}{\partial \lambda_1}
< \frac{{\rm det}(\lambda_1 - X){\rm det}(\lambda_2 - X)}
{{\rm det}(\mu_1 - X){\rm det}(\mu_2 - X)} > |_{\mu_1 =\lambda_1,
\mu_2=\lambda_2}
\ee
whose double discontinuity gives the two-level correlation
function. For this correlation function, the relevant terms in the series $g$
discussed in the previous section,  are those which involve a factor
$(\frac{1}{\tau_{11}} + \frac{1}{\tau_{12}})(\frac{1}{\tau_{23}} +
\frac{1}{\tau_{24}})$
since this factor yields pole terms of the form $\frac{1}{(\lambda_1 -
\mu_1)(\lambda_2 -
\mu_2)}$. Therefore the relevant part of $g$, after taking derivatives and
focusing on terms with poles at $ \mu_i = \lambda_i $, is
\ba\label{gl}
g_L &=& \frac{1}{4N^2} g_t g_u \nonumber\\
&\times& (\frac{1}{i t_1 - u_1} + \frac{1}{i t_1 - u_2} )
(\frac{1}{i t_2 - u_3} + \frac{1}{i t_2 - u_4})
\frac{1}{(\lambda_1 - \mu_1)(\lambda_2 - \mu_2)}\nonumber\\
&\times&(1 - \frac{1}{2N}( \frac{1}{i t_1 - u_3} + \frac{1}{i t_1 - u_4} )
\frac{1}{\lambda_1 - \mu_2})
(1 - \frac{1}{2N}( \frac{1}{i t_2 - u_1} + \frac{1}{i t_2 - u_2}
)\frac{1}{\lambda_2 - \mu_1})
\nonumber\\
\ea
with
\be\label{gt}
g_t = 1 -  \frac{i}{N t_{12}}
\ee
%*****************************************
In the large N limit, we have three types of saddle points.
i) $(t_1^{+},t_2^+,u_1^+,u_2^+,u_3^+,u_4^+)$  or
$(t_1^{-},t_2^-,u_1^-,u_2^-,u_3^-,u_4^-)$
ii) $(t_1^+,t_2^-,u_1^+,u_2^+,u_3^-,u_4^-)$,
$(t_1^+,t_2^-,u_1^-,u_2^-,u_3^+,u_4^+)$,\\
$(t_1^-,t_2^+,u_1^+,u_2^+,u_3^-,u_4^-)$, or
$(t_1^-,t_2^+,u_1^-,u_2^-,u_3^+,u_4^+)$.
iii) $(t_1^+,t_2^-,u_1^+,u_2^-,u_3^+,u_4^-)$, and similar combinations.

 In those expressions the saddle points are  given
\be
   t^{\pm} = \frac{ - i \lambda \pm \sqrt{2 - \lambda^2}}{2}
\ee
\be
   u^{\pm} = \frac{ \mu \pm i \sqrt{2 - \mu^2}}{2}
\ee
with $\lambda_1, \lambda_2, \mu_1,\mu_2$.

For the saddle-points of type  i), $g_u$ is simply
\be
g_u = \sqrt{\vert (u_1 - u_3)(u_1 - u_4)(u_2 - u_3)(u_2 - u_4)\vert }
(\mu_1 -
\mu_2)^2
\ee
Thus we have, as partial contribution of this saddle-point to the correlation
function,
$\rho^{(i)}(\lambda_1,\lambda_2)$ from (\ref{HIZint}) and (\ref{gl}),
\ba\label{94}
  &&\rho^{(i)}(\lambda_1,\lambda_2) = {\rm Im} {\rm \lim_{\mu_1\rightarrow
\lambda_1,
 \mu_2\rightarrow \lambda_2}} \frac{\partial^2}{\partial \lambda_2 \partial
\lambda_1}
  \int dt du \frac{(i t_1)^N (i t_2)^N}{(u_1 u_2 u_3 u_4)^{\frac{N}{2}}}
  \frac{(t_1 - t_2)^2 \prod |u_i - u_j|}{\prod_{j=1}^2\prod_{\alpha=1}^4(i
t_j -
u_\alpha)}
  \nonumber\\
  && (\lambda_1 - \mu_1) (\lambda_1 - \mu_2)^2
  (\lambda_2 - \mu_1)^2 (\lambda_2 - \mu_2)(\frac{1}{i t_1 - u_1} + \frac{1}{i
t_1 - u_2})
  (\frac{1}{i t_2 - u_3} + \frac{1}{i t_2 - u_4})\nonumber\\
  && [ 1 - \frac{i}{N (t_1 - t_2)(\lambda_1 - \lambda_2)}]
  [ 1 - \frac{1}{2N}(\frac{1}{i t_1 - u_3} + \frac{1}{i t_1 -
u_4})\frac{1}{\lambda_1 - \mu_2}]
  \nonumber\\
  &&  [ 1 - \frac{1}{2N}(\frac{1}{i t_2 - u_1} +
  \frac{1}{i t_2 - u_2})\frac{1}{\lambda_2 - \mu_1}]\frac{1}{(\lambda_1 -
\lambda_2)^2}\nonumber\\
  && e^{- N (t_1^2 + t_2^2) - \frac{N}{2}(u_1^2 + u_2^2 + u_3^2 + u_4^2)
  + N(u_1 + u_2)\mu_1 + N(u_3 + u_4) \mu_2 - 2 i N t_1 \lambda_1
   - 2 i N t_2 \lambda_2}\nonumber\\
  &+& {\rm{perm.}}(\lambda_1 \leftrightarrow  \lambda_2, \mu_1
\leftrightarrow  \mu_2)
\ea
from which follows
\ba\label{sin}
  &&\rho^{(i)}(\lambda_1,\lambda_2) = {\rm Im} {\rm \lim_{\mu_1\rightarrow
\lambda_1,
 \mu_2\rightarrow \lambda_2}} \frac{\partial^2}{\partial \lambda_2 \partial
\lambda_1}
  \int dt du \frac{(i t_1)^N (i t_2)^N}{(u_1 u_2 u_3 u_4)^{\frac{N}{2}}}
  \frac{(t_1 - t_2)^2 \prod |u_i - u_j|}{\prod_{j=1}^2\prod_{\alpha=1}^4(i
t_j -
u_\alpha)}
  \nonumber\\
  && (\lambda_1 - \mu_1) (\lambda_1 - \mu_2)
  (\lambda_2 - \mu_1)(\lambda_2 - \mu_2)(\frac{1}{i t_1 - u_1} + \frac{1}{i
t_1
- u_2})
  (\frac{1}{i t_2 - u_3} + \frac{1}{i t_2 - u_4})\nonumber\\
  &&
   \frac{1}{4N^2}(\frac{1}{i t_1 - u_3} + \frac{1}{i t_1 - u_4})
  (\frac{1}{i t_2 - u_1} +
  \frac{1}{i t_2 - u_2})\frac{1}{(\lambda_1 - \lambda_2)^2}\nonumber\\
  && e^{- N (t_1^2 + t_2^2) - \frac{N}{2}(u_1^2 + u_2^2 + u_3^2 + u_4^2)
  + N(u_1 + u_2)\mu_1 + N(u_3 + u_4) \mu_2 - 2i N t_1 \lambda_1 - 2 i N t_2
\lambda_2}
\ea

For those  saddle points of type i),
 the difference $t_1 - t_2$ is of order $1/N$.
Then, cancelling terms between the
numerator  and the
denominator, one  obtains
\ba
\rho^{(i)}(\lambda_1,\lambda_2)&\simeq& {\rm Im} \int dt du
 \frac{(i t_1)^N (i t_2)^N}{(u_1 u_2 u_3 u_4)^{\frac{N}{2}}}
 \frac{|u_1 - u_2||u_3 - u_4|}{(i t_1 - u_1)(i t_1 - u_2)}(\frac{1}{i t_1 -
u_1}
+
 \frac{1}{i t_1 - u_2})\nonumber\\
 && \frac{1}{(i t_2 - u_3)(i t_2 - u_4)}(\frac{1}{i t_2 - u_3} +
 \frac{1}{i t_2 - u_4})\frac{1}{N^2}\nonumber\\
 && e^{- N (t_1^2 + t_2^2) - 2 i N t_1 \lambda_1 - 2 i N t_2 \lambda_2
 - \frac{N}{2}(u_1^2 + u_2^2 + u_3^2 + u_4^2) + N (u_1 + u_2)\lambda_1
 + N (u_3 + u_4)\lambda_2}
 \ea
 which is just a product of  density of states as given
 in the equation (\ref{L11}). Therefore, by considering the saddle point  of
type i)
 $(t_1^{+},t_2^+,u_1^+,u_2^+,u_3^+,u_4^+)$  and
$(t_1^{-},t_2^-,u_1^-,u_2^-,u_3^-,u_4^-)$,
 we obtain
 \be\label{rhorho}
 \rho^{(i)}(\lambda_1,\lambda_2) = \rho(\lambda_1)\rho(\lambda_2)
 \ee
We now return to the integral  (\ref{sin}), and  consider a saddle point of
type ii) ;
one thereby obtains the sin-kernel squared, since now terms like $(i t_1 -
u_1)$ are replaced by the
density of states $\rho(\lambda)$. The exchange of pairs between
$(u_1,u_2)$ and
$(u_3,u_4)$
in the denominator gives
\be\label{rhorhorho}
\rho^{(ii)}(\lambda_1, \lambda_2) = - \frac{\sin^2(\pi N \rho(\lambda)
(\lambda_1 -
\lambda_2))}{N^2
\pi^2 (\lambda_1 - \lambda_2)^2}
\ee
This term adds to (\ref{rhorho}), and by adding both terms, one obtains the
expected vanishing contribution in the limit $\lambda_1 \rightarrow
\lambda_2$.

Now, we take into account the remaining terms, $g_t$ and $g_u$.
In the case ii)  $u_1 - u_3$ is now proportional to the density of
states
$\rho$, which is of order  one.
Then all terms in $g_u$ are of the same order in the Dyson limit, and we have
to return to the discussion of the previous section.
For the saddle point of type iii), such
as $(t_1^+,t_2^-,u_1^+,u_2^+,u_3^-,u_4^-)$, one has
\ba
\rho(\lambda_1,\lambda_2) &=& {\rm Im} \int dt du
\frac{(i t_1)^N (i t_2)^N}{(u_1 u_2 u_3 u_4)^{\frac{N}{2}}}
\frac{2 i t_1 - u_3 - u_4}{(i t_1 - u_3)^2(i t_1 - u_4)^2}
\frac{2 i t_2 - u_3 - u_4}{(i t_2 - u_1)^2(i t_2 - u_2)^2}\nonumber\\
&&[ 1 - \frac{i}{N \pi \rho (\lambda_1 - \lambda_2)} ]
 [ 1 + \frac{i}{4 N \pi \rho (\lambda_1 - \lambda_2)} + \cdots ]\nonumber\\
&\times&  e^{- N (t_1^2 + t_2^2) - 2 i N t_1 \lambda_1 - 2 i N t_2 \lambda_2
 - \frac{N}{2}(u_1^2 + u_2^2 + u_3^2 + u_4^2) + N (u_1 + u_2)\lambda_1
 + N (u_3 + u_4)\lambda_2}
\ea
in which we have replaced various terms by $\rho(\lambda)$ and $(\lambda_1 -
\lambda_2)$
by using their saddle point values. We note that $
\displaystyle t_1^+ - t_2^- = \sqrt{2 -  \lambda^2} = \pi \rho(\lambda)$.
One introduces the same integration variables  $b$ and $\rho$
already used in (\ref{tu}), and integrate over
$\rho_1$ and $\rho_2$
 as in (\ref{rhoint}). Then,  there is a  part of the integrand which has a
double pole in
$\frac{1}{(\lambda_1 -
\lambda_2)^2}$
in the large N limit.
The imaginary parts are then taken for $\lambda_1$ and $\lambda_2$
independently. Thus one obtains  for $g_t$ , after multiplication by a factor
$\displaystyle
 \frac{1}{N \rho(\lambda) (\lambda_1 - \lambda_2)}$,
\ba
&&{\rm Im} \frac{1}{N  \pi \rho (\lambda_1 - \lambda_2)}[ i + \frac{1}{N \pi
\rho
(\lambda_1 - \lambda_2)} ]
e^{- i N \pi \rho (\lambda_1 - \lambda_2)}\nonumber\\
&=& \frac{\cos(\pi N \rho (\lambda_1 - \lambda_2))}{N  \rho(\lambda_1 -
\lambda_2)}
- \frac{\sin(\pi N \rho (\lambda_1 - \lambda_2))}{N^2 \pi^2 \rho^2 (\lambda_1 -
\lambda_2)^2}\nonumber\\
&=&  \frac{d}{d x} \left( \frac{\sin  x}{  x } \right)
\ea
where $x$ stands for $ x=N  \pi \rho (\lambda_1 - \lambda_2)$. Next we return
to the Itzykson-Zuber factor
 $g_u$ of (\ref{A18}). After multiplication  of $g_u$ by  $\displaystyle
 \frac{1}{N \rho(\lambda)(\lambda_1 - \lambda_2)}$ , one obtains
\ba\label{upart}
&&{\rm Im} [\frac{i g_u}{N \rho(\lambda)(\lambda_1 - \lambda_2)} e^{ N
(u_1^- +
u_2^-)\lambda_1
+ N (u_3^+ + u_4^+)\lambda_2 }]\nonumber\\
&=&\frac{\cos  x}{ x} + \frac{\sin  x}{ x^2} - 2 \frac{\cos  x}{
 x^3} - 6
\frac{\sin  x}{ x^4} + \cdots \nonumber\\
&=& \int_{x}^{\infty} \frac{\sin  z}{z} dz
\ea
Therefore, we have by multiplying these two factors,
\be
\rho(\lambda_1,\lambda_2) = - \frac{d}{d x} (\frac{\sin  x}{  x}) \times
\int_{x}^{\infty} \frac{\sin  z}{ z} dz
\ee

The third type of  saddle points iii), for instance
$(t_1^+,t_2^-,u_1^+,u_2^-,u_3^+,u_4^-)$,
does not yield any imaginary part, and may thus be dropped in this  GOE
calculation. However for GSE, they do contribute as well.

Adding the type i) (\ref{rhorho}) and ii) (\ref{rhorhorho}) results, one
obtains the two-point correlation function
for the GOE in the large N limit,
\be
\rho(\lambda_1,\lambda_2) =  \rho^2(\lambda) [1 - (\frac{\sin
x}{ x})^2 -
\frac{d}{d x} ( \frac{\sin  x}{ x}) \times \int_{x}^{\infty}
\frac{\sin  z}{ z} dz ]
\ee
where $x = \pi N \rho(\lambda) (\lambda_1 - \lambda_2)$.
 Comment : This result is of course well-known \cite {Mehta} ; it has been
obtained long ago through the   technique of skew orthogonal polynomials. The
point  of the  long derivation presented here, through generalized
Itzykson-Zuber integrals, is that it  can be repeated in other
cases, such as non-invariant measures involving an external source, for which
the standard method does not apply.

%%%%%%%%%%%%%%%%%%%%%%%%%%%%%%%%%%%%%%%%%%%%%%%%%%%%%%%%%%%
\section{Zero replica limit}

Up to now we have computed averages of ratios  of  characteristic polynomials
defined in (\ref{ratio}). The usual  two-point correlation function of the
resolvent operator is obtained from this ratio by differentiation, as shown in
(\ref{6.1}). Instead of such ratios  one may use an alternative "replica"
method, as follows.

Consider the correlation function
\be\label{10.1}
  F_{2k}(\lambda_1,\lambda_2) = < [{\rm det}(\lambda_1 - X)]^k [{\rm
det}(\lambda_2 - X)]^k >.
\ee
Since $\displaystyle [{\rm det}(\lambda - X)]^k = \exp [k \tr \log (\lambda -
X)]$,
one may recover the correlation functions of the resolvent, by letting   the
 replica number $k$ go to zero :
$k\rightarrow 0$ since :
\be \lim_{k\rightarrow 0} \frac{1}{k^2}
\frac{\partial^2}{\partial \lambda_1 \partial \lambda_2}
F_{2k}(\lambda_1,\lambda_2)
=  <\Tr \frac{1}{\lambda_1 - X} \Tr \frac{1}{\lambda_2 - X} >\ee

Let us first  discuss the zero-replica limit (\ref{10.1}) for  the GUE, and
show that the well-known exact expression for the two-point
correlation function of the resolvent, is  correctly recovered.
There were discussions of the validity of the replica limit in
\cite{KM}.

In an earlier work \cite{BH4},the universality of
$F_{2k}$ , in the Dyson short distance
limit was proven and  its expression was found to be  (for a
probability measure proportional to $\exp{-N Tr V(X)}$) ,
\ba\label{contk}
  F_{2k}(\lambda_1,\lambda_2) &=& e^{\frac{Nk}{2} (V(\lambda_1) +
V(\lambda_2))} e^{- N
k}\frac{1}{k!}
  (2 \pi \rho(\lambda))^{k^2}\nonumber\\
&&\oint \prod_1^k \frac{du_\alpha}{2 \pi} exp - i (
\sum_{\alpha=1}^k u_\alpha)\frac{\Delta^2(u_1,...,u_k)}{\prod_{\alpha=1}^k
(u_\alpha - x)^k (u_\alpha + x)^k}
\ea
a special case   of the general formula  derived  in \cite{BH4}
for different $\lambda_\alpha$'s :
\ba\label{contk2}
&&F_{2k}(\lambda_1,\cdots,\lambda_{2k}) = < \prod_{\alpha=1}^{2k}
{\rm det}(\lambda_\alpha - X) >\nonumber\\
&=&  e^{\frac{N}{2} \sum_{i=1}^{2k} V(\lambda_i)} e^{- N k}\frac{1}{k!}
(2 \pi
\rho(\lambda))^{k^2}
\oint \prod_1^k \frac{du_\alpha}{2 \pi} exp - i (
\sum_{\alpha=1}^k u_\alpha)\frac{\Delta^2(u_1,...,u_k)}{\prod_{\alpha=1}^k
\prod_{l=1}^{2k}
(u_\alpha - x_l)}\nonumber\\
\ea
in which one has used  the scaling variable $x =N \pi
\rho(\lambda)(\lambda_1 -
\lambda_2)$, $x_l = 2 \pi N \rho(\lambda) (\lambda_l - \lambda)$ and
$\lambda=(\lambda_1 + \lambda_2)/2$.
This contour integral   is a compact representation of the sum over the
$(2k)!/(k!)^2$  saddle points which govern the Dyson  limit, and
it is particularly useful for  degenerate cases in which some of the
$\lambda_i$'s are equal.  Indeed in those  cases, subtle
corrections are present in the sum over saddle  points, which are not
easy to handle if one lets the $\lambda$'s approach each other too soon.
The poles due  the degeneracy of
the $\lambda_i$'s have to be cancelled by  the sum over permutations
over all possible saddle points.
The formulae (\ref{contk}) and (\ref{contk2}) do give  a correct answer
for the  degenerate cases of  equal $x_l$'s.

The calculation is done as follows. One first shifts $u_\alpha \to
u_{\alpha}=v_\alpha + x$ and drop all factors in
$F_{2k}$,  which
approach
one in the zero-replica limit. Then we have to deal with the integral
\be
I = \oint  \prod_1^k \frac{dv_\alpha}{2 \pi} \exp (- i k x
-  i \sum v_l )\frac{\Delta^2(v_1,...,v_k)}{ \prod_{1\leq l\leq k} v_l^{2k}
(1 + 2x/v_l)^k}
\ee
which vanishes if one sets $k=0$ in the integrand. The zero-$k$ limit is
slightly tricky since the number of integrations is also equal to $k$, and one
has to find explicit answers for the integrals before approaching the
required limit.  We are interested here  in terms of order
$k^2$ for small $k$.  Thus one expands  the denominator in powers  of $x$,
\be
\frac{1}{(1 + 2 x/v)^k} = 1 -k \sum_1^{\infty} \frac{(-2x)^p}{pv^p} + O(k^2)\ee
Since the one gives a vanishing contour integral, up to order $k^2$ it is
sufficient to expand one of the factors
$(1 + 2 x/v)^{-k}$ to order k, since there are
 $ k$ ways of singling out  a  particular $v_\alpha$ .
If one expanded  both $(1+x/v_\alpha)^{-2k}$ and $(1+x/v_\beta)^{-2k}$, the
number of choices
$k(k-1)/2$, together with the two powers of $k$,
would give a term proportional to $k^3$.

Then , in the zero replica limit, it is sufficient to examine the
integral
\be
I^{(1)}(x) = k  e^{- i k x }\oint \prod
\frac{d v_\alpha}{2 \pi} e^{ - i \sum v_\alpha}
\frac{\Delta^2(v)}{ (1 + 2 x/v_1)^k \prod_1^k v_l^{2k}}
\ee
This integral makes it clear that there are oscillatory terms of the form
$e^{ 2 i x}$, the contribution of the pole $v_1 = - 2 x$, and a
non-oscillating term which comes from the pole $v_1 = 0$.
We will show indeed there are such terms and they are $\displaystyle
(\frac{\sin
x}{x})^2$.
One then expands $e^{-ikx}(v_1+2x)^{-k}$ in  $I^{(1)}(x)$ in powers of  $x$.
This generates the  following integrals,
\ba
\gamma_k^{(p)} &=& \oint \frac{dv}{2 \pi}
\frac{\Delta^2(v) e^{- i \sum v_\alpha}}{v_1^{2 k +  p}v_2^{2k}
\cdots v_{k}^{2k}} \nonumber\\ &=&
(-i)^p \prod_{l=0}^{k-1} \frac{l!}{(k + l)!} \frac{(2 +
p)(4 + p) \cdots ((2 k -  2) + p)(p - 3) (p - 5) \cdots (p - (2 k - 1))}
{(2 k - 1 + p)!}\nonumber\\
&=& (-i)^p \prod_{l=0}^{k-1} \frac{l!}{(k + l)!} \frac{2
\Gamma(2 -
\frac{p}{2})\Gamma(k + \frac{p}{2}) \Gamma(2 k
- p)}{\Gamma(2 k + p)\Gamma(1 + \frac{p}{2})\Gamma(k - \frac{p}{2})\Gamma(3 -
p)}
\ea
Note that when $p$ is an odd integer, $\gamma_k^{(p)}$ does not
contribute to
$I^{(1)}(x)$, since one has to take a real part.  When $p=0$,
$\gamma_k^{(0)}$ coincides with a well-known universal number, related to the
moments  of the Riemann zeta-function. In the zero replica limit of
$k\rightarrow 0$, we have found in an earlier work that \cite{BH5} that
\be
\lim_{k\rightarrow 0}\prod_{l=0}^{k - 1} \frac{l!}{(k + l)!} = 1 + k^2 (1 + c)
+ O(k^3)
\ee
where $c$ is  Euler's constant. Therefore we have
\be
{\lim_{k\rightarrow 0}}\gamma_k^{(p)} = \frac{1}{(1 - p)\Gamma(p + 1)}(-(-i)^p)
\ee
By expanding $\frac{1}{(v_1 + 2 x)^k}$ in  powers  of $x$, we obtain
\be
I^{(1)}(x) = \sum_{p= even}^\infty (- i)^p (\frac{k^2}{p})(2 x)^p
\frac{1}{(1 -
p)\Gamma(p + 1)}
\ee
Thus  the second derivative of $I^{(1)}$ with respect to  $x$ is finally
given by
\ba
\frac{\partial^2 I^{(1)}}{\partial x^2} = - k^2 {\rm Re}\sum_{p=even}^\infty
(- i)^p  \frac{2^p}{p!} x^{p - 2} &=& - k^2 \frac{e^{2 i x} + e^{- 2 i x}
- 2}{2  x^2}\nonumber \\ &=&2k^2 (\frac{\sin x}{x})^2.\ea
Finally there is an additional constant term which comes from the
second derivative of $e^{- i k x}$ in
$I$, which was neglected in $I^{(1)}$.
Therefore adding this constant, we obtain the well known two-point
correlation function of the  GUE by this replica method,
\be
\rho(\lambda_1,\lambda_2) = \rho^2(\lambda) [1 - (\frac{\sin x}{x})^2]
\ee

We may now we proceed to the GOE case.
The Itzykson-Zuber integral for the  GOE case has been discussed in the
previous  section, when we dealt with ratio of characteristic
polynomials. We now consider the following moment
\be
I = < \frac{1}{[{\rm det}(\lambda_1 - X)]^k [{\rm det}(\lambda_2 - X)]^k} >
\ee
Again the zero-replica limit will be used to obtain the
 two-point correlation function of the resolvent operator for the GOE.

Indeed again here we deal again with the  Itzykson-Zuber integral for
a degenerate source, with only  two distinct eigenvalues, $\lambda_1$ and
$\lambda_2$, both
$k$-times degenerate. In such cases the heat-kernel satisfies
the differential equation  (\ref{greplica}).
The asymptotic solution of the solution in the scaling limit,  is given by
(\ref{greplica2}),
\be\label{klimit}
g = 1 + \sum_{p=1}^\infty \frac{[k ( k + 2)( k + 4) \cdots ( k + 2 ( p -
1))]^2}{2^{2 p} p! x^p}
\ee
In the zero-replica limit $k\rightarrow 0$, we need to keep only the
terms proportional to
$k^2$ .  To order $k^2$, we
have from (\ref{klimit}),
\be\label{klimit2}
  g = 1 + \frac{ k^2}{4}  ( \sum_p \frac{(p - 1)!}{p x^p} )
\ee
Then taking the second derivative of $g$with respect to  $x$, we have a
factor
$
\displaystyle (p! + (p - 1)!)/x^{p - 2}$, which is precisely what one
obtains from   the asymptotic expansion of the standard two-point
correlation function of the GOE ensemble ,
\be
\rho_2(x)= \rho^2(\lambda) [1 - (\frac{{\rm sin}x}{x})^2 -
\frac{d}{d x}(\frac{{\rm sin}x}{x})\int_x^{\infty} \frac{{\rm sin}z}{z} dz]
\ee
Indeed taking two derivatives with respect to $x$ of the  second and
third terms :
\be\label{gs}
\frac{1}{2}
\frac{d^2}{d x^2} (\int_x^{\infty} \frac{{\rm sin}z}{z} dz )^2 =
 (\frac{{\rm sin}x}{x})^2 -  \frac{d }{d x}(\frac{{\rm sin}x}{x})
\int_x^{\infty}\frac{{\rm sin}z}{z} dz
\ee
Next let us  compare our expression of $g$ with the factor $\displaystyle
(\int_x^{\infty}
\frac{{\rm sin}z}{z} dz )^2$.
>From the  asymptotic expansion of the integral
\be\label{exp}
\int_x^{\infty}\frac{{\rm sin} z}{z} dz
= {\mathcal {I}}m e^{i x} \sum_{p\geq 0}\frac{p!}{2 x^{p+1}}(- i )^p
\ee
one obtains for  the square of this quantity
\ba\label{exp2}
&&(\int_x^{\infty}\frac{{\rm sin} z}{z} dz)^2 = \frac{1}{2}
\sum_{p,p'}\frac{p!p'!}{x^{p + p'+2} }( - i)^p(i)^{p'}  -\frac{1}{4}e^{2i
x}(\sum
\frac{p!}{2 x^{p+1}}(- i)^{p })^2 \nonumber\\ &&-\frac{1}{4} e^{-2i
x}(\sum
\frac{p!}{2 x^{p+1}}( i)^{p})^2
\ea
Indeed our expansion (\ref{klimit2}) for $g$ in the zero-replica limit
agrees with  the first term,
the non-oscillating term of the above equation,
since
\be
 \sum_{p,p'}
\frac{p!p'!}{x^{p + p'+2}} ( - i)^{p }(i)^{p'} =\int_0^{\infty} \frac{dt}{t}
\e^{-t} \log{(1+ \frac{t^2}{x^2})}
=\sum_{p=0}(-1)^p
\frac{(2p+2)! }{(p+1) x^{2p+2} }
\ee

The Itzykson-Zuber integral which led to (\ref{klimit}), has been derived
dby a saddle-point method  ,  choosing $u_j =
u_+,(j=1,...,k)$ and
$u_l = u_-,(l=k;1,...,2k)$. This gave a
factor $e^{i k x}$ which reduces to unity in the limit $k\rightarrow 0$.
Those saddle-points contribute to the non-oscillating terms. However
there are other saddle-points that one needs to consider in order to
recover the oscillating part. This is analogous to a phenomenon recently
analyzed by Kamenev and M\'ezard \cite {KM}.   As for the  GUE,
the zero-replica limit  requires only
 the $e^{\pm 2 i x}$  oscillating terms.
Those terms   may be
obtained  through
  saddle points of the following type : one divides the $u_i$ (i =
1,...2k) into two groups.
$(u_1,u_2,...,u_k)$ and $(u_{k+1},u_{k+2},...,u_{2k})$.  One chooses one
$u_i$ to
be
a $u_{-}$ in the first group, and the others  $u_{+}$ ; similarly
one $u_{+}$ in the second group,   and the  others are $u_{-}$. For
instance, in the first group,
$u_1 = u_{+}, u_2 = u_{-},u_3=u_{+},...u_{k}=u_{+}$,
and for the second group
$u_{k+1}=u_{+},u_{k+2}=u_{-},....,u_{2k}=u_{-}$.
The combinatorial factor summing over all such choices is
$k^2$. The differential equation for this degenerate case is
similar to (\ref{A17}), but the combinations of $(u_2 - u_j)$
for $j= k+2,...,2k$, and
$(u_i - u_{k+1})$ for $i=1,3,...,k$ are eliminated.
One then modifies  (\ref{degenerate}) to be
\be
f = \sqrt{\vert (u_1 - u_2)(u_3 - u_4)(u_1 - u_3)(u_2 - u_4)\vert }g
\ee
for this purpose, in the  case  $k=2$ given as an exmaple.
Then  $g$ satifies a differential equation which yields the terms
proportional to   $\displaystyle e^{2 i x}$ as a powers series in $1/x$
up to order $1/x^3$.
The first order is
\be
g^{(1)} = \frac{k^2 [ (k-1)^2 - 1]}{4 x}
\ee
and it vanishes in the zero-replica limit.
At second order, one has
\be
g^{(2)} = \frac{9 k^2 [ (k-1)^2 + 1]}{32 x^2}
+ \frac{3 k^2 (k-1)^2 (k - 2)}{16 x^2}
+ \frac{k^2 [ (k-1)^2 (k - 2)^2 - 2 (k - 1)^2]}{32 x^2}
\ee
which come from the contribution of three different
diagrams connecting two lines; a double line,
a connected line, a separate two lines. The diagrams
are the  same as for the non-oscillating calculation, but
the weight factors are different.
In the zero-replica limit, it gives
$g^{(2)} = k^2/4 x^2$.
At third order, we have 7 different diagrams.
Adding thir contributions, one finds in the zero-replica
limit,
\be
g^{(3)} = - \frac{k^2}{2 x^3}
\ee
If one compares this series  with the terms proportional to
$e^{2 i x}$ in (\ref{exp2}),
\be
\frac{k^2}{4} e^{2 i x} ( \sum_p \frac{(p - 1)}{2 x^p}( - i)^{p - 1})^2
= \frac{k^2}{4} e^{2 i x}[ \frac{1}{4 x^2} - \frac{i}{2 x^3} + \cdots]
\ee
one sees that the two agree up to this order.

We have a term $\displaystyle \frac{e^{ i k x}}{x^{k^2}}$.
By the second derivative of this term, we also have $k^2$ term as
\be
   \frac{d^2}{d x^2} \frac{e^{i k x}}{x^{k^2}} = - k^2 ( 1 - \frac{1}{x^2}) +
O(k^3)
 \ee
 We also have similar terms from the derivative of $\displaystyle
 \frac{e^{- i k x + 2 i x}}{x^{k^2}}$. Together, we have
 $\displaystyle 1 - 2 (\frac{\sin x}{x})^2$. Adding this term to
 (\ref{gs}), we obtain the two-point correlation function of GOE,
 \be
  \rho_2(\lambda,\mu) = \rho^2(\lambda) [1 - \left(\frac{\sin x}{x}\right)^2 -
\frac{d}{dx} \frac{\sin x}{x}
  \int_x^\infty \frac{\sin z}{z} dz ]
  \ee
  Thus we have shown here that replica limit of the moment of the
characteristic
polynomial
gives the consistent result with the well known
resolvent two-point correlation functions both for GUE and GOE.

%%%%%%%%%%%%%%%%%%%%%%%%%%%%%%%%%%%%%%%%%%%%%%%%%%%%%%%%%%%%%%%%%%%%%%%%%%%%%%
\section{Gaussian symplectic ensemble}

The Gaussian symplectic ensemble (GSE) is
easily formulated as an extension of  the GOE.
Let $X$ be a quaternion symmetric matrix.
Let us consider as an example the N=2 case ; $X_{11}$ and
$X_{22}$ are both real numbers , whereas  the off-diagonal element $X_{12}$
is a
quaternion and $X_{21}$ its  conjugate.
The quaternion  $X_{12}$ may be written as
\be
X_{12} = a + b e_1 + c e_2 + d e_3
\ee
where $e_i$ are the quaternion basis (i.e. up to a relabelling and a
factor $i$, the Pauli matrices ) ;  in the basis
 $\displaystyle e_1 = \left( \matrix{i&0\cr 0&-i}\right),
e_2= \left(\matrix{0&-1\cr 1&0}\right), e_3= \left(\matrix{0& -i\cr
-i&0}\right)$,  the coefficients
$a,b,c$ and
$d$ are real  numbers. One can write instead the matrix $X$ as
ordinary $4\times 4$ matrix $X'$, the elements are then usual complex
numbers,
\be
X' = \left(\matrix{ x_{11}& 0 & u & i v\cr
      0& x_{11}& i v^* & u^*\cr
      u^* & - i v & x_{22} & 0\cr
       - i v^* & u & 0 & x_{22} }\right)
\ee
where $x_{11}$ and $x_{22}$ are real,
$u$ and $v$  complex.
The relation
$[{\rm det}(\lambda - X)]^2 = {\rm det}( \lambda - X')$,
aloows us to write
\be\label{F}
F_2(\lambda,\mu) = < \frac{[{\rm det}(\lambda - X)]^2}{[{\rm det}(\mu -
X)]^2} >
= < \frac{{\rm det}(\lambda - X')}{{\rm det}(\mu - X')} >.
\ee
The density of state $\rho(\lambda)$
may be easily deduced  from $F_2$. From the relation
(\ref{F}), one obtains in the  N=2 case,
\be
F_2(\lambda,\mu)
=  \int \oint \frac{(\lambda - t_1)^2 (\lambda - t_2)^2}{(\mu - t_1)^2
(\mu - t_2)^2} (t_1 - t_2)^4 e^{- (t_1^2 + t_2^2)} \frac{dt_1}{2\pi}
\frac{ dt_2}{2 \pi}
\ee
in which $t_1$ and $t_2$ are the eigenvalues of $X$.
The imaginary part of $F_2$, from which one deduces the density of
eigenvalues through
\be
\rho(\lambda) = \frac{1}{\pi N}\lim_{\mu\rightarrow \lambda} {\rm Im} \frac{
\partial}{\partial \lambda}
F_2(\lambda,\mu)
\ee is obtained by picking up the contribution of the pole
$t_1 = \mu$. Taking the imaginary part, a derivative with respect to
$\lambda$ and setting
$\mu\rightarrow \lambda$
afterwards, one obtains
\be
\rho(\lambda) = \int (\lambda - t_2)^4 e^{- (\lambda^2 + t_2^2)}
\frac{dt_2}{2 \pi}
\ee
which may easily be checked  directly.

For the inverse of the characteristic polynomial,
we find easily the following formula in the $X'$ representation (still
for the $N=2$ example),
\ba
< \frac{1}{{\rm det}(\mu - X')} > &=& \int  \prod_{i=1}^4dz_i dz_i^* dX'
e^{  i 2 z_a^* ( \mu \delta_{ab} - X'_{ab}) z_b -  \tr
X'^2}\nonumber\\
&=& \int\prod_{i=1}^4dz_i dz_i^* e^{ -2 (\sum z_i^* z_i)^2 + i 2 \mu z_a^*
z_a}
\nonumber\\
&=& \frac{\pi^2}{4}\int db \prod_{i=1}^4dz_i dz_i^* e^{-{2} b^2 - 2i b
\sum z_i^* z_i + 2i
 \sum \mu z_i^* z_i
 }
\nonumber\\
& =& \int db \frac{1}{(\mu - b)^4} e^{- 2 b^2}
\ea
The imaginary part is  the Hermite polynomial $H_3(\mu)$.

For general N, similarly, we find that the expectation value of the
inverse of the
  characteristic polynomial is
\be
< \frac{1}{{\rm det}(\mu - X')} > =
 \int db \frac{1}{(\mu - b)^{2N}} e^{- {N} b^2}
\ee
and its imaginary part is simply $H_{2N - 1}(\mu)$.

For  the inverse of the product of
two characteristic polynomials, one writes
\be\label{nu1}
< \frac{1}{({\rm det}(\mu_1 - X))^2 ({\rm det}(\mu_2 - X))^2} >
= \int \frac{1}{[{\rm det}(B)]^{2N}} e^{- \frac{N}{2} {\rm tr} B^2 + N {\rm
tr} B M}
dB
\ee
where $X$ is an $N \times N$ quaternion symmetric matrix,
$B$ a  $2 \times 2$ quaternion matrix, and $M$ is a diagonal
matrix $M = diag(\mu_1,\mu_2)$.

The average of the square of the characteristic polynomials may then be
written as
\be\label{nu2}
< [{\rm det}(\lambda - X)]^2 >
= \int dA [{\rm det}( \lambda - A ) ]^N e^{- \frac{N}{2} {\rm tr} A^2}
\ee
where $A$ is a $2 \times 2$ real symmetrix matrix. The quantity $[{\rm
det}( \lambda - A ) ]^N$ is a polynomial in $\lambda$ of order
$\lambda^{2N}$.

Finally  the ratio $F_N(\lambda,\mu)$ may be written as an integral over
a super-matrix $Q$ ,  the derivation being similar to that for  GOE,
\be
  F_N(\lambda,\mu) = \int \frac{1}{({\rm Sdet} Q )^N} e^{- N {\rm Str} Q^2 +
  i N {\rm Str} Q \Lambda}
  dQ
\ee
A super-group  diagonalization and the Itzykson-Zuber integral
( for which we may  use the same formulae as for  GOE since the
Jacobian has the same form  after the diagonalization) leads then to
\be
F_N(\lambda,\mu) = \int dt_1 dt_2 du \frac{(t_1 t_2)^N}{u^{2N}}
\frac{|t_1 - t_2|(\lambda - \mu)^2}{(u - it_1)(u - i t_2)}
[ 1 - \frac{1}{N(\lambda - \mu)}(
\frac{1}{u - it_1 } + \frac{1}{u - i t_2} )]
\ee
The density of state $\rho(\lambda)$ follows :
\ba
\rho(\lambda) &=& < \frac{1}{N}{\rm Tr} \delta (\lambda - X) >\nonumber\\
&=& \lim_{\mu\rightarrow \lambda} {\rm Im} \int_{-\infty}^{\infty}
 dt_1 dt_2 du \frac{(t_1 t_2)^N}{u^{2N}}
  \frac{|t_1 - t_2| (2 u - it_1 - it_2)}{(u - i t_1 )^2 (u - i t_2)^2}
  \nonumber\\
  && e^{- \frac{N}{2}(t_1^2 + t_2^2) - N u^2 - iN(t_1 + t_2)\lambda
  - 2 i N u \mu}\nonumber
  \ea
  The formula is quite similar to that of the  GOE, except that the
combination
   $\displaystyle \frac{(t_1 t_2) }{b^{2}}$ is raised here
  to the power  $2N$ instead of $-N$. This difference makes the
calculation of the imaginary part easier, since
   the contour integral on $b$ gives a contribution  to the imaginary
part,
  similar to that of GUE, and therefore the  result does not involve
the incomplete Gaussian  integrals (such as
  $B(x)$ in (\ref{B})),
  which appear in the GOE case.

  Let us consider now a  ratio of  characteristic polynomials
$F_N(\lambda_1,\lambda_2,
  \mu_1,\mu_2)$
  defined as
  \be
  F_N(\lambda_1,\lambda_2, \mu_1,\mu_2)
  = < \frac{[{\rm det}(\lambda_1 - X) {\rm det}
  (\lambda_2 - X)]^2}{[{\rm det}(\mu_1 - X) {\rm det}
  (\mu_2 - X)]^2} >
  \ee
  With the supermatrix formalism and its supergroup diagonalization,
   one can write again an integral over eigenvalues, similar to
(\ref{FN2}) in the GOE  case,
  \ba
  F_N(\lambda_1,\lambda_2, \mu_1,\mu_2)
  &=& \int \frac{(t_1 t_2 t_3 t_4)^N}{(u_1 u_2)^{2N}}
  \frac{\prod_{i<j}^4
  |t_i - t_j| (u_1 - u_2)^4}{\prod_{\alpha=1}^2 \prod_{k=1}^4
  (u_\alpha - i t_k)} e^{- \frac{N}{2} \sum t_i^2 - N \sum u_\alpha^2}
  \nonumber\\
  &\times& I dt_1 dt_2 dt_3 dt_4 du_1 du_2
  \ea
  where $I$ is the HIZ integral,
  \be
    I = \int dg e^{ N {\rm Str } g \hat Q g^{-1} \Lambda}
    \ee
   with $\hat Q = diag( it_1,it_2,it_3,it_4,u_1,u_1,u_2,u_2)$,
   and $\Lambda =
diag(\lambda_1,\lambda_1,\lambda_2,\lambda_2,\mu_1,\mu_1,\mu_2,\mu_2)$.

   The HIZ integral is in fact identical to that for the GOE case
with corresponding  variables. In the supergroup integration for the GOE
case one found a product of two series, one infinite, and the other one
finishing after a finite number of terms. The same structure appears in
the present supergroup integration for the GSE case, except that the
finite and infinite series have their resepctive variables switched.

    Thus in the large N limit, one has to consider, as in section 8,
three types of saddle-points ;  the saddle points of class i) and ii) give
the same answer.
    The saddle points for $t_i$ and $u_\alpha$ become
    \be
    t_{\pm} = \frac{- i \lambda \pm \sqrt{4 - \lambda^2}}{2}
    \hskip 5mm u_{\pm} = \frac{ \mu \pm i \sqrt{4 - \mu^2}}{2}
    \ee
    Note that for the present GSE case, the density of state $\rho(\lambda)$
    is given by $\rho(\lambda) = \sqrt{4 - \lambda^2}/2\pi$.
    Therefore, by the same arguments as for the  GOE, we obtain the
two-point correlation  function

    \be\label{gse1}
    \rho_2(\lambda_1,\lambda_2) = \rho^2(\lambda) [
    1 - (\frac{\sin 2x}{2 x})^2 - \frac{d}{dx} (\frac{\sin 2x}{2x})
\int_x^\infty
    \frac{\sin 2z}{2 z} dz ]
    \ee

    We have neglected the third class of the saddle point iii) for the
GOE. In the GSE,  those
    saddle points  do contributite, since the imaginary part is taken
from a contour integral over  $b_1$ and $b_2$. Since the HIZ formula for
$\beta=4,k=1$
    takes  a simple form, we obtainfor the saddle point values
   of those b's  $(b_1^+,b_2^-)$, and thus
    \be\label{gse2}
       I = \frac{d}{d x} \frac{\sin 2x}{2x}
     \ee
     For the corresponding $t$ integral, we take the following saddle
points :
     a set of saddle points such as $(t_1^+,t_2^-,t_3^+,t_4^-)$  gives
     the non-oscillating constant contribution. Note that
$t_1^+ t_2^- = 1$, and
     $i N (t_1^+ + t_2^-) \lambda_1 = - \lambda^2 N$.
Thus there is a correction
to
     (\ref{gse1}) by (\ref{gse2}). By taking the normalized coefficient,
which
     makes the two-point correlation function becomes $O(x^4)$ in the
small $x$  limit
     , a property of the  GSE, we find the scaling limit of the
two-point  correlation
     function,
     \ba\label{GSEtwoP}
     \rho_2(\lambda_1,\lambda_2) &=& \rho^2(\lambda) \left(
     1 - (\frac{\sin 2x}{2 x})^2 - \frac{d}{dx} (\frac{\sin 2x}{2x})
[\int_x^\infty
    \frac{\sin 2z}{2 z} dz - \frac{\pi}{2} ]\right)\nonumber\\
    &=&\rho^2(\lambda) \left(
     1 - (\frac{\sin 2x}{2 x})^2 + (\frac{d}{dx} \frac{\sin 2x}{2x}) \int_0^x
    \frac{\sin 2z}{2 z} dz  \right)
    \ea

%************************************************************

\section{Extension to an external matrix source}

Let us consider an external matrix source $A$  be coupled to a real
symmetric random matrix, or to a quaternion self dual random matrix. The
corresponding Gaussian probability measure
\be P_A(X) = \frac{1}{Z} \e^{-\frac{N}{2}{\rm tr} X^2 + N {\rm tr}A X}\ee
has lost the invariance under GOE or GSE.

 In a previous article \cite{BH3}, we have considered this external
source problem for the correlation functions  $\langle\prod_{\alpha=1}^k {\rm
det}(\lambda_\alpha - X)\rangle$, in which  $X$ is a real symmetric random
matrix. This is done by  integrating over $X$
\be
\int e^{- \frac{N}{2}{\rm tr} X^2 + N {\rm tr}A X + i N {\rm tr} X Y} dX
= e^{- \frac{N}{4} {\rm tr}[ (Y - i A)^2 + (Y - i A)(Y^T - i A)]}
\ee
where $Y = - \sum_{\alpha = 1}^k \bar \theta_{ \alpha a} \theta_{ \alpha b}$.
One may assume, without loss of generality, that $A$ is a diagonal
matrix.  Then,  the only new term with respect to the zero-source case,
is  ${\rm exp}[ -i N \sum a_j \bar \theta_{ \alpha j} \theta_{ \alpha
j}]$. This diagonal  term modifies the previous determinant $({\rm det}
B)^N$, and gives instead
$\prod_{j=1}^N \prod_{l=1}^k (t_l - i a_j)$. In this way we had obtained
in \cite{BH4} that, when all the
$\lambda_j$ are equal to $\lambda$, one has
\be
\langle [{\det}(\lambda -X)]^k \rangle  = e^{- N \sum \lambda_l^2} \int
\prod_{l=1}^k
\prod_{j=1}^N
(t_l - i a_j) \prod_{l<l'}(t_l - t_{l'})^4 e^{-N\sum t_l^2 + 2 i N \lambda
\sum
t_l}
\prod_{l=1}^k d t_l.
\ee

    Similarly for the  ratio  of two characteristic
polynomials, we have
 $Y = - \sum_{\alpha=1}^k \bar \theta_{\alpha a} \theta_{\alpha b} -
 z_{\alpha a}^* z_{\alpha b}$ as in (\ref{Y}).
 The external source gives a diagonal shift for $\theta$ and $z$.
 This leads to the modification of the Sdet term.
 For instance, in the k=1 case, one replaces the
 super-determinant $[{\rm Sdet}]^{\frac{N}{2}}$ by
 \be
 \frac{1}{[{\rm Sdet} Q]^{\frac{N}{2}}} \rightarrow
 \prod_{\gamma = 1}^N
 \frac{(a_\gamma - i t)}{[(a_\gamma - u_1)(a_\gamma -
u_2)]^{\frac{1}{2}}}
 .\ee
Then the density of states, for the GOE modified by an external source
matrix,
 is given by
  \ba\label{L111}
    \rho(\lambda)
    &=& - \frac{1}{8 \pi^2 N} {\rm Im} \lim_{\mu \rightarrow \lambda}
    \int dt du \prod_{\gamma = 1}^N \frac{(a_\gamma -i t)}{[(a_\gamma -
u_1)(a_\gamma - u_2)]^{\frac{1}{2}} }
    \frac{|u_1 - u_2|}{(i t - u_1) (i t - u_2)}\nonumber\\
     &\times&[ \frac{1}{it - u_1} + \frac{1}{i t - u_2}]
     e^{- 2 i N t \lambda  + N(u_1 + u_2)\mu - N t^2 -
\frac{N}{2} (u_1^2 + u_2^2)}
     \ea

 %%%%%%%%%%%%%%%

Let us consider the one point Green function $G(z)$ (with an external
source) ,
\be
G(z) = < {\rm tr} \frac{1}{z - X} >
\ee
for a real symmetric random matrix $X$.
It is given by the derivative of $F(\lambda,\mu)$,
\ba
G(z) &=& {\lim_{\lambda = \mu = z}}\frac{\partial}{\partial \lambda}
<\frac{{\rm
det}(\lambda - X)}{{\rm det}(\mu - X)} >
\nonumber\\
&=& \int \prod_{\gamma=1}^N \frac{a_\gamma - i t}{(u_1 - a_\gamma)(u_2 -
a_\gamma)}\frac{|u_1 - u_2|
(2it - u_1 - u_2)}{(it - u_1)^2
(it - u_2)^2}\nonumber\\
&\times & e^{- 2 i N t \lambda  + N(u_1 + u_2)\mu - N t^2 -
\frac{N}{2} (u_1^2 + u_2^2)} dt du_1 du_2
\ea
Using the change of variables, $b= \frac{u_1 + u_2}{2}$, $r = \frac{(u_1 -
u_2)^2}{4}$,
we obtain
\ba
G(z) &=& \int dt \int db \int_0^\infty dr
\frac{
(it - b)}{((it)^2 - 2 i t b + b^2 - r)^2}
 e^{- 2 i N t \lambda  + 2 N b\mu - N t^2 -
N (b^2 + r)} \nonumber\\ &\times &\prod_{\gamma} \frac{a_\gamma - i
t}{(b^2 - r - 2 a_\gamma b +  {a_\gamma}^2)^{1/2}}
\ea
After  integration  by parts over r, we obtain an expression,  similar
to the GUE case,
\be\label{G1}
G(z) = \int \prod_\gamma (1 - \frac{it}{N(b - a_\gamma)})
\frac{1}{it} e^{- \frac{1}{2 N} t^2 - i t b}\frac{db}{2 \pi i} \frac{dt}{2 \pi}
\ee
In the large N  limit, one may make the replacement
\be\label{G2}
\prod_{\gamma = 1}^N ( 1 - \frac{it}{N (b - a_\gamma)} )
\simeq \exp ( -\frac{it}{N} \sum_{\gamma=1}^N \frac{1}{b - a_\gamma} ).\ee
Let us denote the (non-random)  density of states of the external
matrix $A$ as $\rho_0(a)$, we may then write the r.h.s. of (\ref{G2})
as
\be
\exp ( - it \int da \frac{\rho_0(a)}{u - a} ).
\ee
We define the resolvent of the external source $G_0(z)$
\be
G_0(z)
= \int da \frac{\rho_0(a)}{z- a}
\ee
>From (\ref{G1}) and (\ref{G2}), we obtain
\be
\frac{\partial G}{\partial z}
= \oint \frac{db}{2 \pi i} \frac{1}{u + G_0(u) - z}
\ee
 The
contour surrounds all the eigenvalues $a_\gamma$.
The zeros of the denominator satisfy
\be\label{saddleu}
u + \frac{1}{N}\sum \frac{1}{u - a_\gamma} = z
\ee
As discussed in a previous paper \cite{BH1}, we take
the poles of $\hat u(z) = z - \frac{1}{z} + O(\frac{1}{z^2})$ and
$u=\infty$, and then
\be
\frac{\partial G}{\partial z} = 1 - \frac{1}{1 + \frac{d G_0}{d \hat u(z)}}
= 1
- \frac{d \hat u(z)}{d z}
\ee
The integration gives
\be
G(z) = z - \hat u(z)
\ee
Since
$\hat u(z)$ is a solution of $u + G_0(u) = z$, we obtain the following
 equation, due to Pastur \cite{Pastur} for the GUE
\be
G(z) = G_0(z - G(z))
\ee
Thus we obtain the same Pastur equation as for the GUE case ;
this is easily understandable from a diagrammatic analysis ;  in the large
N limit,  planar diagrams are simple rainbow diagrams and do not
distinguish between  GOE, GSE  or   GUE at leading order in the
large N limit. For  the GSE, a similar  algebra would lead to
the same equation.

For the two-point correlation function , k=2, the same shift for the ${\rm
Sdet}$ occurs in  presence of the external source.
We have found in the previous sections that the resolvent two point
correlation functions in GOE and GSE
 in the Dyson scaling limit. We now discuss the
two point correlation function in the Dyson
scaling limit when the external source matrix is coupled to the
random matrix. We have already given a proof of the universality in GUE.
The argument goes as follows for this  GUE case \cite{BH1}.

In the presence of a matrix source there is a  kernel
$K_N(\lambda_1,\lambda_2)$  given by
\be\label{kernelGUE}
K_N(\lambda_1,\lambda_2) = \int \frac{dt}{2 \pi}\oint \frac{du}{2\pi i}
\frac{1}{i t} \prod_{\gamma=1}^N ( 1 + \frac{it}{N(u - a_\gamma)}) e^{
-\frac{t^2}{2N} - i u t - i t \lambda_1 + N u (\lambda_1 - \lambda_2)},
\ee
from which all the n-point correlation functions may be obtained by the usual
determinant formulae of a matrix whose elements are the
$K_N(\lambda_i,\lambda_j)$. Defining the scaling variable
$y = N (\lambda_1 -
\lambda_2)$, and using, in the large N limit, the expression of  (\ref{G2})
for  the product $ \prod_{\gamma=1}^N ( 1 + \frac{it}{N(u - a_\gamma)}) $,
after  integration over $t$, one obtains
\be
\frac{\partial K_N}{\partial \lambda_1}
= \frac{1}{\pi} {\rm Im} \oint \frac{du}{2 \pi i} \frac{1}{
u + G_0(u) - \lambda_1 + i \epsilon} e^{-u y}.
\ee
Again one defines the pole $\hat u$, and
obtain
\ba
\frac{\partial K_N}{\partial \lambda_1}
&=& \frac{1}{\pi} {\rm Im} \frac{d \hat u}{d \lambda_1} e^{- y \hat u
(\lambda_1
- i \epsilon)}
\nonumber\\
&=& - \frac{1}{\pi y}\frac{\partial}{\partial \lambda_1} {\rm Im}
 ( e^{- y \hat u (\lambda_1 - i \epsilon)})
\ea
in which
\be
\hat u (\lambda_1 - i \epsilon) = \lambda_1 - {\rm Re} G(\lambda_1) - i \pi
\rho(\lambda_1).
\ee
Therefore one ends up with
\be
K_N(\lambda_1,\lambda_2) = - \frac{1}{\pi y} e^{- y [\lambda_1 - {\rm Re}
G(\lambda_1)]}
\sin [ \pi y \rho(\lambda_1) ].
\ee
Putting this expression of the kernel into the correlation functions, one
sees that their only dependence in the eigenvalues $a_\gamma$ of the external
source  is a  scale factor through the density of state ; for intsnace the
connected tow-point function is simply
\be
\rho_c(\lambda_1,\lambda_2) = - \frac{1}{\pi^2 y^2} \sin^2 [ \pi
\rho(\frac{\lambda_1 + \lambda_2}{2}) y ].
\ee

  We now  consider the universality in the GOE and GSE ensembles in the
presence of an
  external source.
  In the GOE case, we used a generalized  HIZ integral which had two parts. One
part was the
  t-integration, which gave simply
  \be
  g_t = 1 - \frac{i}{N (t_1 - t_2) (\lambda_1- \lambda_2)}
  \ee
   This leads now to the integral
   \be
   \int \prod_\gamma^N (a_\gamma - i t_1)(a_\gamma - i t_2)
\frac{g_t}{\lambda_1 - \lambda_2}
   e^{- N t_1^2 - N t_2^2 + 2 i N t_1 \lambda_1 + 2 i N t_2 \lambda_2}
\frac{dt_1}{2 \pi}
   \frac{d t_2}{2 \pi}
   \ee
    In the large N limit, the saddle point equations for the $t_j (j=1,2)$ are

    \be
    t_j - i \lambda_1 - \frac{1}{2 N}\sum_\gamma \frac{1}{a_\gamma - i t_j} = 0
     \ee
     As in (\ref{saddleu}), there is a saddle point $\hat t_j$, which
behaves as
     $\hat t_j \simeq i \lambda_j$ in the large $\lambda_j$ domain.
    Then $\hat t_j$ becomes
     \be
     \hat t_j = i \lambda_j - i {\rm Re} G(\lambda_j) + \pi \rho(\lambda_j)
     \ee
     Thus, this t-integration in the large N limit behaves as in the
sourceless case, and yields  $\displaystyle \frac{d}{dx} ( \frac{\sin
x}{x})$
     where $x$ is $N \pi \rho[(\lambda_1 + \lambda_2)/2] (\lambda_1 -
\lambda_2)$.
    More genrally, this part may be written as
    \be\label{dkernel}
    I = \frac{d}{d x} K(x,y)
    \ee
    where $K(x,y)$ is a kernel. This is due to the HIZ integral for $\beta=4$.
    In our case, this kernel is a sine-kernel.

    For the u-integration, one has a product
    \be
    \prod_\gamma^N \prod_{j=1}^4 \frac{1}{u_j - a_\gamma}
    = \exp[ - \sum_\gamma \sum_j \ln (u_j - a_\gamma) ]
    \ee
    There is a saddle point $\hat u_j$, which behaves as
    $\hat u_1 = \hat u_2 \simeq \lambda_1$ and $\hat u_3 = \hat u_4 \simeq
\lambda_2$.
    Using these saddle points, the factor coming from this u-integral
is the same as in(\ref{upart}).
    If we denote this u-integral by I,
    it is easy to find that the derivative of I with respect to x is
    the sine kernel. More generally, we have
    \be
    \frac{d }{d x} I = K(x,y)
    \ee
    This is due to HIZ formula for $\beta = 1$ which gives
    \be
    \sum \frac{n!}{(i x)^{n+1}}
    \ee
    In the sine kernel case, we have
    \be
    \frac{d }{d x} [ e^{i x}( \frac{1}{x} - \frac{i}{x^2} - \frac{2}{x^3} +
\cdots )]
    = i \frac{e^{i x}}{x}
    \ee
    where the successive cancellation occurs in the higher order.

     Thus we have a universal two point correlation function in the presense
of  the
     external source,
     \be
\rho(\lambda_1,\lambda_2) =  \rho^2(\lambda) [1 - (\frac{\sin
x}{ x})^2 -
\frac{d}{d x} ( \frac{\sin  x}{ x})  \int_{x}^{\infty}
\frac{\sin  z}{ z} dz ]
\ee
where $x = \pi N \rho(\lambda) (\lambda_1 - \lambda_2)$ and the external source
$a_\gamma$ appears only in the density of state $\rho$. The GSE case may be
analyzed   as the GOE
and yields also  universal  correlation function with respect to the external
source eigenvalues, as  in (\ref{GSEtwoP}).

\section{Universalities at the edges in GOE}
 Near  edges of the support of the density of states, it is well-known that a
new scaling behavior takes place (\cite {{BB},{TW1}}. In the simplest case of
the Wigner semi-circle   the behavior is governed by an
  Airy kernel. Let us consider now the edge behavior of
 the two point correlation of  characteristic polynomials,
 \be
 F_N(\lambda_1,\lambda_2) = < {\rm det}(\lambda_1 - X) {\rm det}(\lambda_2
- X)
>
 \ee
 when  $X$ is a real symmetric random matrix.
 If $\lambda_1$ and $\lambda_2$ are within the bulk of the support of  the
asymptotic density of states, we  have
 found in the Dyson scaling limit that \cite{BH3}
 \be
 F_N(\lambda_1,\lambda_2) = \frac{1}{x} \frac{d}{dx} \frac{\sin x}{x}
 \ee
 As we have discussed earlier , by using Grassmann variables, one finds for
arbitrary $\lambda_1,\lambda_2$ and finite $N$,
 \ba\label{Ai}
 &&F_N(\lambda_1,\lambda_2) = \int ({\rm det}B)^N e^{- N \tr B^2 + i N \tr B
\Lambda}
 \nonumber\\
 &=& \int (t_1 t_2)^N e^{- N t_1^2 - N t_2^2 - 2 i N t_1 \lambda_1 - 2 i N t_2
\lambda_2}
 [ \frac{(t_1 - t_2)^2}{(\lambda_1 - \lambda_2)^2} - \frac{i (t_1 -
t_2)}{(\lambda_1 -
 \lambda_2)^3}] dt_1 dt_2
 \ea
 In the large N limit, the saddle points for the  $t_a$'s are
 \be
 t_a = \frac{- i \lambda_a \pm \sqrt{2 - \lambda_a^2}}{2}
 \ee
 The critical point corresponds to a degenerate quadratic form of
fluctuations near the saddle point ; expanding then to next order (since there
is a flat direction), one finds a new scaling limit when the  parameters
$\alpha$ and
$\beta$ are
 $\alpha= \frac{2}{3}$ and $\beta = \frac{1}{3}$, when the $\lambda$'s are at
distance $N^{-\alpha}$ of the end point $\sqrt 2$, and the $t_a$'s at
distance $N^{-\beta}$ of $-i/\sqrt2$.
Performing   the scaling change  of variables
 \ba
 \lambda_a &=& \sqrt{2} - N^{-\alpha} x_a \nonumber\\
 t_a &=& - \frac{i}{\sqrt{2}} + N^{-\beta} \tau_a.
 \ea

  the integral becomes, in this regime  of large $N$,
 \ba
 F^{(1)}(\lambda_1,\lambda_2)
 &=& \int e^{- \frac{2 \sqrt{2}}{3} i (\tau_1^3 + \tau_2^3) + 2 i (x_1
\tau_1 +
x_2 \tau_2)}
 [ \frac{(\tau_1 - \tau_2)^2}{(\lambda_1 - \lambda_2)^2}
 + \frac{(\tau_1 - \tau_2)}{(\lambda_1 - \lambda_2)^3}] d \tau_1 d \tau_2
 \nonumber\\
 &=& \frac{1}{(\lambda_1 - \lambda_2)^2} [ Ai''(x_1) Ai(x_2) - 2 Ai'(x_1)
Ai'(x_2)
 + Ai(x_1) Ai''(x_2)]\nonumber\\
 \ea
 in which use has been made of the differential
equation satisfied by the Airy function,
$Ai''(x) = x  Ai(x)$.
 Noting that
 \ba
 &&(\frac{\partial}{\partial x_1} - \frac{\partial}{\partial x_2}) [ Ai'(x_1)
Ai(x_2)
 - Ai'(x_2) Ai(x_1)] \nonumber\\
 &=& - [  Ai''(x_1) Ai(x_2) - 2 Ai'(x_1) Ai'(x_2)
 + Ai(x_1) Ai''(x_2)]
 \ea
  the second term of (\ref{Ai}) is  then simply  the Airy kernel divided
  by $(\lambda_1 - \lambda_2)$.
  One   ends up with
  \be
  F_N(\lambda_1,\lambda_2) =
  \frac{1}{x_1 - x_2}(\frac{\partial}{\partial x_1} - \frac{\partial}{\partial
x_2})
  [\frac{ Ai'(x_1) Ai(x_2)
 - Ai'(x_2) Ai(x_1)}{x_1 - x_2}]
 \ee
 This  confirms the result   (\ref{dkernel}), which was stated for more
general  kernels.

 Extending this analysis  to  ratios of  characteristic polynomials,
 one obtains similarly an  other term,   expressible as  an
integral of the kernel.
 Thus we have obtained the two point correlation function near the edge
 for the  GOE ensemble. The same argument can easily be transposed to the
GSE.

 Let us now  consider other edge problems that one meets when an  external
source matrix is added to the probability distribution.
 For instance one can tune  the external source to create a gap in the
spectral density of states $\rho(\lambda)$. At the critical point
 at which this gap closes, a new universality  class appears and we
have studied earlier the new scaling behavior at the origin \cite{BH7,BH8}.

It was found that the  kernel $K_N(x,y)$ for the  GUE ensemble in the
appropriate scaling limit, was
\be
 K(x,y) = \frac{\hat \phi'(x) \hat \psi'(y) - \hat \phi''(x) \hat \psi(y) -
\hat \phi(x) \hat
 \psi''(y)}{x - y}
 \ee

For the  GOE case, in the critical domain of  the gap closing point, one finds
a two point correlation function,
\be
\rho(x,y) = 1 - K(x,y)K(y,x) - \frac{d}{dx} K(x,y) \int_x^\infty K(z,y) dz
\ee

\section{Level spacing distribution in GOE and GSE}

The level spacing probability function $E(s)$, the probability
 that there is no eigenvalue inside the interval $[-s/2,s/2]$,
 is given by the Fredholm determinant
 \ba
 E(s) &=& {\rm det}[ 1 - \hat K ]\nonumber\\
 &=& \sum_{n=0}^{\infty} \frac{(-1)^n}{n!} \int_{-\frac{s}{2}}^{\frac{s}{2}}
 \cdots \int_{-\frac{s}{2}}^{\frac{s}{2}}{\rm det}[ K(x_i,x_j)]_{i,j=1,...,n}
 \prod_{k=1}^n d x_k
 \ea
 Let us briefly review the derivation of this formula for $E(s)$
\cite{TW,BH8}. This level spacing function is expressed by the Hamiltonian
formalism. The derivation is due to Tracy and Widom \cite{TW}.

In the GUE case,
we have
\be
E(a,b) = {\rm det}[ 1 - \hat K]
\ee
where we choose the interval $(a,b)$ in general, and later
we put $(a,b) = (- \frac{s}{2},\frac{s}{2})$.
The kernel $K(x,y)$ acts on this interval, and we define
\be
\hat K(x,y) = K(x,y) \theta(y - a) \theta(b - y)
\ee
in which $\theta(x)$ is the Heaviside function.
By  definition, the derivative of $E$ with respect to the end points is
\be
\frac{\partial \ln E(a,b)}{\partial b} = - \tilde K(b,b)
\ee
where
\be
\tilde K = \frac{\hat K}{1 - \hat K}
\ee
 ,$\tilde K$ is the Fredholm resolvant.
One has a similar derivative with respect to $a$. Then, one obtains
\be
\frac{d E(s)}{ds} = \frac{1}{2} (\frac{\partial}{\partial b} -
\frac{\partial}{\partial a})\ln E(s)_{b=-a= s/2} = - \tilde
K(\frac{s}{2},\frac{s}{2})
\ee
This leads to
\be
E(s) = \exp[ - \int_0^s H(\frac{z}{2},\frac{z}{2}) dz ].
\ee
where we write $H(x,x) = \tilde K(x,x)$.
For GUE, the kernel is the sine kernel,
\be
K(x,y) = \frac{\phi(x)\phi'(y) - \phi'(x) \phi(y)}{x - y}.
\ee
Operating with $K$ on $\phi$, one obtains
\be
q(x) = <x|\frac{1}{1 - \hat K} | \phi>,
p(x) = <\phi'| \frac{1}{1 - \hat K}|x>.
\ee
This leads to
\be
\tilde K(x,y) = \frac{q(x)p(y) - q(y) p(x)}{x - y}.
\ee
Taking the  derivatives of $q(x)$ and $p(x)$ with respect to $b$
, one obtains  a set of equations, which are a  Hamiltonian system,
\be
\dot Q = P(1 - \frac{2 Q^2}{b}),
\dot P = Q (\frac{2 P^2}{b} - 1)
\ee
where $Q(b) = q(b,-b:b)$ and $P(b) = p(b,-b:b)$,
$q(b,-b,:b)$ , obtained by setting $a=-b,x=b$
in $q(b,a:x)$, (We have written explicitly the
interval dependence of
$q(x)$).
The Hamiltonian $H$ which governs this dynamical system is simply
\be\label{Hamiltonian}
H(b,b) = P^2 + Q^2 - \frac{2 P^2 Q^2}{b}
\ee
($P$ and $Q$ have Hamiltonian form, $\dot Q = \frac{\partial H}{\partial P}$,
 $\dot P = - \frac{\partial H}{\partial Q}$, with $H = \tilde K(b,b)$).
For small $s$, we obtain from these two equations,
\be
P = 1 + s + \frac{7}{8}s^2 + \frac{65}{72} s^3 + \cdots
\ee
\be
Q = \frac{s}{2} - \frac{s^3}{48} + \cdots
\ee
This leads to
\be
E(s) = 1 - s  + O(s^4)
\ee
which agrees with the well known result \cite{Mehta}.

For the  GOE ensemble, the kernel is a quaternion matrix, and the correlation
functions are expressed by a quaternion determinant.

The matrix kernel of GOE, $\sigma(x,y)$, is
\be
\sigma(x,y) = \left( \matrix{ s(x,y) & D s(x,y)\cr
     J s(x,y)& s(x,y) }\right)
\ee
where we denote the sine kernel as $s(x,y) = \sin (x - y)/(x - y)$, and
\be
J s(x) = \int_0^x s(y) dy - \epsilon(x) = \epsilon s(x) - \epsilon
\ee
where $\epsilon(x) = \frac{1}{2} {\rm sgn} x$.

 Tracy and Widom have found the Fredholm determinant
for this matrix kernel \cite{TW}. Their result for the interval $(a,b)$
reads
\be\label{GOEES}
E(s) = \exp[ - \frac{1}{2}\int_0^s H(z,z)dz - \frac{1}{2}\int_0^s H(z,-z)dz ]
\ee
for GOE. The Hamiltonian $H$ is same as the Hamiltonian for the  GUE given in
(\ref{Hamiltonian}).
We have
\be
H(b,-b) = \frac{Q(b) P(b)}{b}
\ee
 For  small $s$, one obtains from (\ref{GOEES}),
\be
E(s) = 1 - s + \frac{1}{36} s^3 + O(s^4)
\ee
which agrees with the known result \cite{Mehta} (We have dropped a
factor $\pi$ in the sine kernel. The correct coefficient is
$\frac{\pi^2}{36}$).

For GSE, we have
\be\label{GSEES}
E(\frac{s}{2}) = \frac{1}{2} [ \exp[- \frac{1}{2} \int_0^s (H(z,z) +
H(z,-z))dz]
+  \exp[- \frac{1}{2} \int_0^s (H(z,z) - H(z,-z))dz] ]
\ee
For  small $s$, it gives
\be
E(s) = 1 - s + O(s^4)
\ee

 When the external source matrix $A$ has only two dsitinct eigenvalues -a and
+a, each of them  $
\frac{N}{2}$ times degenerate,  a gap in the spectrum around the origin
may be created
 by  tuning appropriately the parameter $a$,  .
 At some critical value of $a$ the gap closes and its vicinity leads to a new
interesting universality class. For the GUE case,  we have given  in an
earlier work \cite{BH8} the equation satisfied by $E(s)$in the oppropriate
scaling vicinity of  the gap closing point.
 In this gap closing case, the problem is again governed by a Hamiltonian
system  with now  three different
 $Q_i,P_i (i = 0,1,2)$, as shown  in \cite{BH8}.
 They satisfy
\be
\dot Q_n= \frac{\partial H}{\partial P_n}
\ee
\be
\dot P_n = - \frac{\partial H}{\partial Q_n}
\ee
The Hamiltonian $H = \tilde K(b,b)$ reads
\ba
H &=& b P_2 Q_0 + Q_2 P_1 + Q_1 P_0 - u P_1 Q_0 - v P_2 Q_1\nonumber\\
&+& \frac{1}{b - a} [ P_1(b) Q_1(a) - Q_2(b) P_2(a) - Q_0(b) P_0(a)]\nonumber\\
&\times& [P_1(a) Q_1(b) - Q_2(a) P_2(b) - Q_0(a) P_0(b)]
\ea
for the interval $(a,b)$.

The kernel $\tilde K(b,-b)$ is
 \be
 \tilde K(b,-b) = \frac{Q_1(b) P_1(-b) - Q_2(b) P_2(-b) - Q_0(b) P_0(-b)}{2 b}
 \ee
 Using the expressions of (\ref{GOEES}) and (\ref{GSEES}),
we obtain the level spacing function both for GOE and GSE
near the gap closing point.
 The second derivative of $E(s)$ is the level spacing probability $p(s)$,
which is the probability density that two successive eigenvalues lie at
 distance $s$.
For $s$ small, it is easy to verify that $p(s)$ is  linear
 in $s$, a characteristic behavior of GOE. The small $s$ expansion may be
computed from the
expansions of $Q_n,P_n$, which are
$\displaystyle
Q_0(b) = \frac{\sqrt{2}}{4 \pi} \Gamma(\frac{1}{4}) +
\frac{\sqrt{2}}{2\pi^{3/2}}b + O(b^2),
Q_1(b) = - \frac{\sqrt{2}}{2 \pi} \Gamma(\frac{3}{4}) b + O(b^2),
Q_2(b) = - \frac{\sqrt{2}}{2 \pi} \Gamma(\frac{3}{4}) + O(b),
P_0(b) =  - \frac{1}{3\sqrt{\pi}}b^3 + O(b^4),
P_1(b) = - \frac{1}{\sqrt{\pi}} + O(b),
P_2(b) = \frac{1}{\sqrt{\pi}} b + O(b^2).$

Thus we have obtained  new results for the level spacing probablity
of GOE.
 This may be  important,  for instance,
in the discussions
of  universality for the the energy spectrum of
 quantum dots with
 interactions \cite{Boris}. Our study of the external source problem
may be related to the questions of
 the distribution of cycles in the permutations with
 external source
 \cite{BK} in the GOE case, and to the crystal growth in a random environment
\cite{GTW} or
 to the spin glass problem \cite{Parisi}.

\section{General $\beta$ }

 We have discussed   hereabove the  HIZ integration for arbitrary $\beta$.
The two point correlations of the characteristic polynomials are given for the
general $\beta$ case by

\be
< {\rm det}(\lambda_1 - X){\rm det}(\lambda_2-X) >
=
C \int \prod_{i=1}^N(\lambda_1 - x_i) \prod_{j=1}^N (\lambda_2 - x_j)
\prod_{i<j} |x_i - x_j|^\beta
e^{- N \sum x_i^2 } \prod_{i=1}^N dx_i
\ee
in which  the eigenvalues of $X$ are $x_i$.
In the $\beta = 1$ GOE case,
this correlation function is simply
\be
F(\lambda_1,\lambda_2) = - \sqrt{\frac{\pi}{2 x^3}} J_{\frac{3}{2}}(x) =
\frac{1}{x} \frac{d}{dx} (\frac{\sin x}{x})
\ee
In the GUE case, this two point correlation function
becomes the kernel itself \cite{BH4} $K_N(\lambda_1,\lambda_2)$, and it is
given by
$\displaystyle
\sqrt{\frac{\pi}{ 2 x}} J_{\frac{1}{2}}(x) =
\frac{\sin x}{x}$.

It is easy to derive the general expressions for arbitrary $\beta$
from the relation (\ref{A29}). If we  use the dual $\beta$, which is equal
to $\frac{4}{\beta}$ and $k=1$ in (\ref{A29}) ???.

The HIZ integral has a finite number terms when
$\frac{2}{\beta} - \frac{1}{2} $ is a half integer ; for instance for
$\beta=1$ , the GOE ensemble is dual to the GSE, for which we know that the
expansion is finite . From the saddle-point large N analysis,   this two point
correlation function of  characteristic polynomials
 in the Dyson limit is
\be
F =  \sqrt{\frac{\pi}{2}} \frac{1}{x^{\frac{2}{\beta}- \frac{1}{2}}}
  J_{\frac{2}{\beta} - \frac{1}{2}} (x)
\ee
The  Bessel functions of half-integer order is expressed by
\be
 J_{n + \frac{1}{2}}(x) =\sqrt{\frac{2}{\pi}} x^{n + \frac{1}{2}} (-\frac{d}{x
dx})^n (\frac{\sin x}{x}).
\ee
 For  instance,
 when $\beta = 4$, $k=1$, (GSE), we have shown that indeed it is a Bessel
function
 $J_0(x)$ in (\ref{A11}).

The expression in terms of  Bessel functions has been derived for the
generalized Selberg integrals by Aomoto
\cite{Aomoto}, and further studied by Forrester \cite{Forrester}. It may be
interesting to note the following fact.

\section{Summary}

In this paper, we have studied the correlations
of the characteristic polynomials of random matrices,
which are either  real symmetrix (GOE) or quaternionic self-dual
(GSE). It was shown  that they are universal
in the Dyson's limit with respect to an  external matrix source linearly
coupled to  the random matrix. As usual the correlation functions
are only sensitive to
the external source through a scale set by the mean spacing.

For the ratio of  characteristic polynomials,
we have applied  supersymmetric techniques, and
obtained as a by-product the resolvent correlation functions of the GOE and
GSE ensembles. The method required using group integrals which are no longer of
the semi-classical type studied by Harish-Chandra, Itzykson and Zuber.
However we have found that, in the limit required by Dyson scaling, these
integrals can be performed both for the  GOE and the GSE. We have also used  an
alternative method, in which one computes ratios of characteristic polynomials,
and  we have obtained again the  correlation functions  by a   replica
method, in the zero-replica limit.

The level spacing probability has been  studied
for the  GOE and the  GSE ensemble, and we have given an explicit
representation for E(s) in the closing gap case,
 obtained by  tuning of the external source.

\vskip5mm

{\bf Acknowledgement}

S.H. has benefited from  a Grant-in-Aid for Scientific Research (B) 
of JSPS.

%%%%%%%%%%%%%%%%%%%%%%%%%%%%%%%%%%%%%%%%%%%%

\end{document}